\documentclass[10pt, journal, compsoc, final]{IEEEtran}

\pdfoutput=1

\usepackage{subfigure}

\usepackage{url}
\usepackage{amstext,amssymb,amsfonts, amsmath}
\usepackage{algorithmic, algorithm}
\usepackage[autostyle]{csquotes}  
\usepackage{mathtools, cuted}
\usepackage{array}

\usepackage[colorinlistoftodos,textsize=tiny,shadow]{todonotes}
\usepackage{marginnote}
\makeatletter
\renewcommand{\@todonotes@drawMarginNoteWithLine}{%
\begin{tikzpicture}[remember picture, overlay, baseline=-0.75ex]%
    \node [coordinate] (inText) {};%
\end{tikzpicture}%
\marginnote[{
    \@todonotes@drawMarginNote%
    \@todonotes@drawLineToLeftMargin%
}]{
    \@todonotes@drawMarginNote%
    \@todonotes@drawLineToRightMargin%
}%
}
\makeatother

\usepackage{authblk}
\usepackage[autostyle]{csquotes}  

\usepackage[backend=bibtex,style=ieee,doi=false, isbn=false, url=false, eprint=false, maxnames=6]{biblatex}
\newcommand{\R}{\mathbb{R}}

\newcommand{\bs}[1]{\boldsymbol{#1}}
\newcommand{\Vector}[1]{\bs{#1}}
\newcommand{\Matrix}[1]{\mathbf{#1}}

\newcommand{\Loss}{\bs{\mathcal{L}}}
\newcommand{\Reg}{\bs{\mathcal{R}}}

\DeclareMathOperator*{\argmin}{\bs{\arg}\!\bs{\min}}

\begin{document}


\title{A Critical Survey of Deconvolution Methods for Separating cell-types  in Complex Tissues}

\author[1]{Shahin Mohammadi$^{*, }$\thanks{$*$ Corresponding authors: mohammadi@purdue.edu, ayg@cs.purdue.edu}}
\author[2]{Neta Zuckerman$^{\star, }$\thanks{$\star$ Currently at: Genentech Inc., South San Francisco, CA 94080, USA}}
\author[2]{Andrea Goldsmith}
\author[1]{ Ananth Grama$^{*, }$}
\affil[1]{Department of Computer Sciences, Purdue University, West Lafayette, IN 47907, USA}
\affil[2]{Department of Electrical Engineering, Stanford University, Stanford, CA 94305, USA}

%

\markboth{Foundations \& Applications of Science of Information - Special Issue of Proceedings of IEEE}%
{Shell \MakeLowercase{\textit{et al.}}: Bare Demo of IEEEtran.cls for Journals}
%


\maketitle

\begin{abstract} 
Identifying properties and concentrations of components from an observed mixture,
known as \textit{deconvolution}, is a fundamental problem in signal processing. It has diverse 
applications in fields ranging from hyperspectral imaging to denoising readings from biomedical sensors. 
This paper focuses on \textit{in-silico} deconvolution of signals associated with complex tissues into
their constitutive cell-type specific components, along with a quantitative characterization of the
cell-types. Deconvolving mixed tissues/cell-types is useful in the removal of contaminants (e.g.,
surrounding cells) from tumor biopsies, as well as in monitoring
changes in the cell population in response to treatment or infection. In these contexts, the observed
signal from the mixture of cell-types is assumed to be a convolution, using a linear
instantaneous (LI) mixing process, of the expression levels of genes in constitutive
cell-types. The goal is to use known signals corresponding to individual cell-types
along with a model of the mixing process to cast the deconvolution problem as a suitable
optimization problem.

In this paper, we present a survey and in-depth analysis of models, methods, and assumptions
underlying deconvolution techniques. We investigate the choice of the different loss functions
for evaluating estimation error, constraints on solutions, preprocessing and data filtering,
feature selection, and regularization to enhance the quality of solutions, along with the
impact of these choices on the performance of commonly used regression-based methods for
deconvolution. We assess different combinations of these factors and use detailed statistical
measures to evaluate their effectiveness. Some of these combinations have been proposed in
the literature, whereas others represent novel algorithmic choices for deconvolution. We identify
shortcomings of current methods and avenues for further investigation. For many of the identified
shortcomings, such as normalization issues and data filtering, we provide new solutions. We summarize our 
findings in a prescriptive step-by-step process, which can be applied to a wide range of deconvolution problems.

\end{abstract}


\markboth{Special Issue of Proceedings of IEEE - Foundations \& Applications of Science of Information,~Vol.~X, No.~X, X~2015}%
{Mohammadi \MakeLowercase{\textit{et al.}}: Deconvolution Methods}


\IEEEpeerreviewmaketitle

\section{Introduction}

\IEEEPARstart{S}{ource} separation, or deconvolution, is the problem of 
estimating individual signal components from their mixtures. This problem arises when source signals are 
transmitted through a mixing channel and the mixed sensor readings are observed.
Source separation has applications in a variety of fields.
One of its early applications was in processing audio signals 
\cite{Pedersen2004, Meng2012, Emmanuel2014, Labastie2015}. Here, mixtures of different 
sound sources, such as speech or music, are recorded simultaneously using several microphones. 
Various frequencies are convolved by the impulse response of the room and the goal is to separate 
one or several sources from this mixture. This has direct applications in speech 
enhancement, voice removal, and noise cancellation in recordings from populated 
areas. In hyperspectral imaging, the spectral signature of each pixel is observed.
This signal is the combination of pure spectral signatures of constitutive elements mixed according to their relative 
abundance. In satellite imaging, each pixel represents sensor readings for different patches 
of land at multiple wavelengths. Individual sources correspond to reflectances 
of materials at different wavelengths that are mixed according to the material 
composition of each pixel \cite{Gillis2014, Ma2014, Nuzillard2000, Pauca2006, Villeneuve2012}.

Beyond these domains, deconvolution has applications in denoising biomedical sensors.
Tracing electrical current in the brain is widely used as a proxy for 
spatiotemporal patterns of brain activity. These patterns have significant clinical 
applications in diagnosis and prediction of epileptic seizures, as well as characterizing 
different stages of sleep in patients with sleep disorders. Electroencephalography (EEG) 
and magnetoencephalography (MEG) are two of the most commonly used techniques for cerebral 
imaging. These techniques measure voltage fluctuations and changes in the electromagnetic fields,
respectively. Superconducting QUantum Interference Device (SQUID) sensors used in 
the latter technology are susceptible to magnetic coupling due to geometry and must be 
shielded carefully against magnetic noise. Deconvolution techniques are used to separate
different noise sources and ameliorate the effect of electrical and magnetic 
coupling in these devices \cite{Tang2000, Hesse2006, Vaya2007, Zhang2012}.

At a high level, mixing channels can be classified as follows: (i) 
linear or nonlinear, (ii) instantaneous, delayed, or convolutive, and (iii) over/under 
determined. When neither the sources nor the mixing process is available, the problem is known 
as blind source separation (BSS). This problem is highly under-determined in general, and 
additional constraints; such as independence among sources, sparsity, or non-negativity; are 
typically enforced on the sources in practical applications. On the other hand, a new class of 
methods has been developed recently, which is known as semi or guided BSS \cite{Pedersen2004, 
Hesse2006, Emmanuel2014, Labastie2015}. In these methods, additional information is available 
a priori on the approximate behavior of either sources or the mixing process. In this paper, 
we focus on the class of over-determined, linear instantaneous (LI) mixing processes, for 
which a deformed prior on sources is available. In this case, the parameters of the linear 
mixer, as well as the true identity of the original sources are to be determined.

In the context of molecular biology, deconvolution methods have been used to identify
constituent cell-types in a tissue, along with their relative proportions.
The inherent heterogeneity of tissue samples makes it difficult to identify separated,
cell-type specific signatures, i.e., the precise gene expression levels for each cell-type.
Relative changes in cell proportions, combined with variations attributed to the changes
in the biological conditions, such as disease state, complicate identification of
true biological signals from mere technical variations. Changes in tissue composition are
often indicative of disease progression or drug response. For example, coupled depletion
of specific neuronal cells with the gradual increase in the glial cell population is
indicative of neurodegenerative disorders. An increasing proportion of malignant
cells, as well as a growing fraction of tumor infiltrating lymphocytes (TIL) compared
to surrounding cells, directly influence tumor growth, metastasis, and clinical outcomes
for patients \cite{Kuhn2012, Newman2015}. Deconvolving tissue biopsies allows further
investigation of the interaction between tumor and micro-environmental cells, along with
its role in the progression of cancer.


The expression level of genes, which is a proxy for the number of present copies of each
gene product, is one of the most common source factors used for separating cell-types and
tissues. In the linear mixing model, the expression of each gene in a complex mixture is
estimated as a linear combination of the expression of the same gene in the constitutive
cell-types. \textit{In silico} deconvolution methods for separating complex tissues can be
coarsely classified as either \textit{full deconvolution}, in which both cell-type specific
expressions and the percentages of each cell-type are estimated,
or \textit{partial deconvolution} methods, where one of these data sources is used to
compute the other \cite{Shen-Orr2013}. These two classes loosely relate to BSS and
guided-BSS problems. Note that in cases where relative abundances are used to estimate
cell-type-specific expressions, the problem is highly under-determined, whereas in the
complementary case of computing percentages from noisy expressions of purified cells,
it is highly over-determined. The challenge is to select the most reliable features that
satisfy the linearity assumption. We provide an in-depth review of the recent deconvolution
methods in Section~\ref{sec:overview}.

In contrast to computational methods, a variety of experimental cell separation techniques have 
been proposed to enrich samples for cell-types of interest. However, these experimental methods
not only involve significant time, effort, and expense, but may also result in insufficient 
RNA abundance for further quantification of gene expression. In this case, amplification
steps may introduce technical artifacts into the gene expression data. Furthermore, sorting
of cell-types must be embedded in the experiment design for the desired subset of cells,
and any subsequent separation is infeasible. Computational methods, on the other hand,
are capable of sorting mixtures at different levels of resolution and for arbitrary
cell-type subsets of interest.


The organization of the remainder of the paper is as follows: in Section~\ref{sec:biology} we introduce
the basic terminology from biology needed to formally define the deconvolution problem in
the context of quantifying cell-type fractions in complex tissues. The formal definition of
the deconvolution problem and its relationship to linear regression is defined in
Section~\ref{sec:decon_def}. Sections~\ref{sec:obj_choice} and \ref{sec:obj_examples}
review different choices and examples of the objective function used in regression.
An overview of computational methods for biological deconvolution is provided in
Section~\ref{sec:overview}. Datasets and evaluation measures used in this study are
described in Sections \ref{sec:datasets} and \ref{sec:eval_measures}, respectively.
The effect of the loss function, constraint enforcement, range filtering, and feature
selection choices on the performance of deconvolution methods is evaluated systematically in
Sections~\ref{sec:loss_effect}-\ref{sec:regularization_effect}.
Finally, a summary of our findings is provided in Section~\ref{sec:conclusion}.

\section{Background and Notation} 
\label{sec:notations}

\subsection{Biological Underpinnings}
\label{sec:biology}

The \textit{central dogma} of biology describes the flow of genetic information within
cells -- the genetic code, represented in DNA molecules,
is first transcribed to an intermediate construct, called \textit{messenger RNA (mRNA)},
which in turn translates into \textit{proteins}. These proteins are the functional workhorses of the cell.
Genes, defined as the minimal coding sections of the DNA, contain the recipe for
making proteins. These instructions are utilized dynamically by the cell to adapt to
different conditions. The amounts of various proteins in a cell can be measured at a time
point. This corresponds to the level of \textit{protein expression}. 
This process is limited by the availability of high-quality antibodies that can specifically
target each protein. The amount of active \textit{mRNA} in a cell, however,
can be measured at the genome scale using high-throughput technologies such as \textit{microarrays} and
\textit{RNASeq}. The former is an older technology that relies on the binding affinity
of complementary base pairs (alphabets used in the DNA/RNA molecules), while
the latter is a newer technique, using \textit{next generation sequencing (NGS)}.
This technique estimates gene expression based on the overlap of mRNA fragments with
known genomic features. Since microarrays have been used for years, extensive databases
from different studies are publicly available. RNASeq datasets, in comparison,
are relatively smaller but growing rapidly in scale and coverage. 
Both of these technologies provide reliable proxies for the amount of proteins in cells,
with RNASeq being more sensitive, especially for lowly expressed genes. A drawback of
these methods, however, is that the true protein expression is also regulated by
additional mechanisms, such as post-transcriptional modifications, which cannot be
assayed at the mRNA level.

The expression level of genes is tightly regulated in different stages of cellular
development and in response to environmental changes. In addition to these
\textit{biological variations} due to cellular state,
intermediate steps in each technology introduce \textit{technical variations} in
repeated measurement of gene expression in the same cell-type. To enhance reproducibly
of measurements, one normally includes multiple instances of the same cell-type in each
experiment, known as \textit{technical} replicates. The expression profiles from these
experiments provide a snapshot of the cell under different conditions. In addition
to biological variation of genes within the same cell-type, there is an additional
level of variation when we look across different cell-types. Some genes are
ubiquitously expressed in all cell-types to perform housekeeping functions,
whereas other genes exhibit \textit{specificity} or \textit{selectivity} for one,
or a group of cell-types, respectively. A compendium of expression profiles of
different cells at different developmental stages is the data substrate for
\textit{in silico} deconvolution of complex tissues.

\subsection{Deconvolution: Formal Definition}
\label{sec:decon_def}

We introduce formalisms and notation used in discussing 
different aspects of \textit{in silico} deconvolution of biological signals. We focus on 
models that assume \textit{linearity}, that is, the expression signature of the mixture 
is a weighted sum of the expression profile for its constitutive cell-types. 
In this case, sources are cell-type specific references and the mixing process is 
determined by the relative fraction of cell-types in the mixture.

We first introduce the mathematical constructs used:

\begin{itemize}

\item $\Matrix{M} \in \R^{n \times p}$: Mixture matrix, where each entry $\Matrix{M}(i, j)$ 
represents the raw expression of gene $i$, $1 \leq i \leq n$, in heterogeneous sample
$j$, $1 \leq j \leq p$. Each sample, represented by $\Vector{m}$, is a column of the matrix
$\Matrix{M}$, and is a combination of gene expression profiles from constituting cell
types in the mixture.

\item $\Matrix{H} \in \R^{n \times r}$: Reference signature matrix for the expression of primary cell 
types, with multiple biological/technical replicates for each cell-type. In this matrix, rows 
correspond to the same set of genes as in $\Matrix{M}$, columns represent replicates
and there is an underlying grouping among columns that collects profiles 
corresponding to the same cell-type.

\item $\Matrix{G} \in \R^{n \times q}$: Reference expression profile, where the expression
of similar cell-types in matrix $\Matrix{H}$ is represented by the average value.

\item $\Matrix{C} \in \R^{q \times p}$: Relative proportions of each cell-type in the mixture 
sample. Here, rows correspond to cell-types and columns represent samples in mixture 
matrix $\Matrix{M}$.

\end{itemize}

Using this notation, we can formally define deconvolution as an optimization problem that 
seeks to identify ``optimal" estimates for matrices $\Matrix{G}$ and $\Matrix{C}$,
denoted by $\hat{\Matrix{G}}$ and $\hat{\Matrix{C}}$, respectively. 
Since $\Matrix{G}$ and/or $\Matrix{C}$ are not known a priori, we use an 
approximation that is based on the linearity assumption. In this case, we aim to find 
$\hat{\Matrix{G}}$ and $\hat{\Matrix{C}}$ such that their product is close to 
the mixture matrix, $\Matrix{M}$. Specifically, given a function $\bs{\delta}$
that measures the distance between the true and approximated solutions, also
referred to as the loss function, we aim to solve:

\begin{equation}
\min_{0 \leq \hat{\Matrix{G}}, \hat{\Matrix{C}}} \bs{\delta}(\hat{\Matrix{G}}\hat{\Matrix{C}}, \Matrix{M}) 
\label{eq:decon_mat}
\end{equation}

In partial deconvolution, either $\Matrix{C}$ or $\Matrix{G}$, or their noisy representation,
is known a priori and the goal is to find the other unknown matrix. When matrix $\Matrix{G}$,
referred to as the \textit{reference profile}, is known, the problem is
over-determined and we seek to distinguish features (genes) 
that closely conform to the linearity assumption, from the rest of the (variable) genes.
In this case, we can solve the problem individually for each mixture sample. Let us
denote by $\Vector{m}$ and $\hat{\Vector{c}}$ the expression profile and estimated
cell-type proportion of a mixture sample, respectively. Then, we can rewrite
Equation~\ref{eq:decon_mat} as:

\begin{equation}
\min_{0 \leq \hat{\Vector{c}}} \bs{\delta}(\Matrix{G}\hat{\Vector{c}}, \Vector{m})
\label{eq:decon_vec}
\end{equation}
 
This formulation is essentially a linear regression problem, with an arbitrary loss function. 
On the other hand, in the case of full deconvolution, we can still estimate $\Matrix{C}$ in a 
column-by-column fashion. However, estimating $\Matrix{G}$ is highly under-determined and we 
must use additional sources to restrict the search space. One such source of information is 
the variation across samples in $\Matrix{M}$, depending on the cell-type concentrations in the 
latest estimated value of $\Matrix{C}$. In general, most regression-based methods for full 
deconvolution use an iterative scheme that starts from either noisy estimates of $\Matrix{G}$ 
and $\Matrix{C}$, or a random sample that satisfies given constraints on these matrices, 
and successively improves over this initial approximation. This iterative process can be 
formalized as follows:

\begin{eqnarray}
\label{eq:ANLS}
\hat{\Matrix{C}} &\leftarrow& \argmin\limits_{0 \leq \hat{\Matrix{C}}} (\bs{\delta}(\hat{\Matrix{G}}\hat{\Matrix{C}} - \Matrix{M})) \\ \nonumber
\hat{\Matrix{G}} &\leftarrow& (\argmin\limits_{0 \leq \hat{\Matrix{G}}} (\bs{\delta}(\hat{\Matrix{C}}^T\hat{\Matrix{G}}^T - \Matrix{M})^T) )^T\\ \nonumber
\end{eqnarray}

Please note that the updating $\hat{\Matrix{G}}$ is typically row-wise (for each gene), 
whereas updation of $\hat{\Matrix{C}}$ is column-wise (for each sample). Non-negative matrix
factorization (NMF) is a dimension reduction technique that aims to factor each column
of the given input matrix as a nonnegative weighted sum of non-negative basis vectors,
with the number of basis vectors being equal or less than the number of columns in the
original matrix. The \textit{alternating non-negative least squares formulation (ANLS)}
for solving NMF can be formulated using the framework introduced in Equation~\ref{eq:ANLS}.
There are additional techniques for solving NMF, including the \textit{multiplicative updating
rule} and the \textit{hierarchical alternating least squares (HALS)} methods, all of which
are special cases of block-coordinate descent \cite{Kim2013}. Two of the most common
loss functions used in NMF are the Frobenius and Kullback-Leibler (KL)
divergence \cite{Kim2013}.

In addition to \textit{non-negativity (NN)}, an additional \textit{sum-to-one (STO)} constraint
is typically applied over columns of the matrix $\hat{\Matrix{C}}$, or the sample-specific vector
$\hat{\Vector{c}}$. This constraint restricts the search space, which can potentially enhance the 
accuracy of the results, and  simplifies the interpretation of values in $\hat{\Vector{c}}$ as relative percentages. Finally, another
fundamental assumption that is mostly neglected in prior work is the \textbf{similar cell
quantity (SCQ)} constraint. The similar cell quantity
assumption states that all reference profiles and corresponding mixtures must be normalized
to ensure that they represent the expression level of the ``same number of cells." If
this constraint is not satisfied, differences in the cell-type counts directly affect
concentrations by rescaling the estimated coefficients to adjust for the difference.


In this paper, we focus on different loss functions ($\bs{\delta}$ functions), as
well as the role of constraint enforcement strategies, in estimating $\hat{\Vector{c}}$.
These constitute the key building blocks of both partial and full deconvolution methods.

\subsection{Choice of Objective Function}
\label{sec:obj_choice}

In linear regression, often a slightly different notation is used, which we describe here. 
We subsequently relate it to the deconvolution problem. Given a set of 
samples, $\{ (\Vector{x}_i, y_i) \}_{i=1}^m$, where $\Vector{x}_i \in \R^k$ and $y_i \in \R$, 
the regression problem seeks to find a function $f(\Vector{x})$ that minimizes the aggregate error over all samples.
Let us denote the fitting error by $r_i = y_i - f(\Vector{x}_i)$. Using this notation, 
we can write the regression problem as:

\begin{equation}
\argmin_{f \in \mathcal{F}} \sum_{i = 1}^m \Loss (r_i)
\label{eq:reg_formulation}
\end{equation}

\noindent where the \textit{loss function} $\Loss$ measures the cost of estimation error. 
We focus on the class of linear functions, that is $f_{\Vector{w}} (\Vector{x}) = 
\Vector{w}^T\Vector{x}$, for which we have $r_i = y_i - \Vector{w}^T\Vector{x}_i$. In 
this formulation, $y_i$ corresponds to the expression level of a gene in the mixture,
vector $\Vector{x}_i$ is the expression level of the same gene in the reference cell
types, and $\Vector{w}$ is the fraction of each cell-type in the mixture.
We can represent $\{\Vector{x}_i\}_{i=1}^m$ in the compact form by matrix $\Matrix{X}$, in which row $i$ corresponds to $\Vector{x}_i$.

In cases where the number of parameters is greater than the number of samples, that is matrix $\Matrix{X}$ is a \textit{fat matrix}, minimizing Equation~\ref{eq:reg_formulation}, directly, can result in the \textit{over-fitting} problem. Furthermore, when features (columns of $\Matrix{X}$) are highly correlated, solution may change drastically in response to small changes in the samples, specifically among the correlated features. This condition, known as \textit{multicollinearity}, can result in inaccurate estimates, in which coefficients of similar features are greatly different. To remedy these problems, we can add a
\textit{regularization term}, which incorporates additional constraints
(such as sparsity or flatness) to enhance the stability of results. We re-write
the problem with the added regularizer as:

\begin{equation}
\argmin_{\Vector{w} \in \R^k} \{ \underbrace{\sum_{i = 1}^m \Loss (y_i  - \Vector{w}^T\Vector{x}_i)}_{\mbox{Overall loss}} + \underbrace{\lambda \Reg (\Vector{w})}_{\mbox{Regularizer}} \}
\label{eq:linReg}
\end{equation}

\noindent where the $\lambda$ parameter controls the relative importance of estimation error 
versus regularization. There are different choices and combinations for the loss function 
$\Loss$ and regularizer function $\Reg$, which we describe in the following sections.

\subsubsection{Choice of Loss Functions}
\label{sec:loss}

There are a variety of options for suitable loss functions. Some of these 
functions are known to be asymptotically optimal for a given noise density, whereas 
others may yield better performance in practice when assumptions underlying the
noise model are violated. We summarize the most commonly used set of 
loss functions:

\begin{itemize} 

\item If we assume that the underlying model is perturbed by Gaussian white noise, 
the squared or quadratic loss, denoted by $\Loss_{2}$, is known to be 
asymptotically optimal. This loss function is used in classical least squares regression 
and is defined as:

\begin{equation*}
\Loss_{2}(r_i) = r_i^2 = (y_i - \Vector{w}^T\Vector{x}_i)^2
\end{equation*}


\item Absolute deviation loss, denoted by $\Loss_{1}$, is the optimal choice if noise
follows a Laplacian distribution. Formally, it is defined as:

\begin{equation*}
\Loss_{1}(r_i)=|r_i| = |y_i - \Vector{w}^T\Vector{x}_i|
\end{equation*}

Compared to $\Loss_2$, the choice of $\Loss_1$ is preferred in the presence of outliers,
as it is less sensitive to extreme values


\item Huber's loss function, denoted by $\Loss_{huber}^{(M)}$, is a parametrized 
combination of $\Loss_1$ and $\Loss_2$. The main idea is that $\Loss_2$ loss is more 
susceptible to outliers, while it is more sensitive to small estimation errors. To 
combine the best of these two functions, we can define a half-length parameter $M$, 
which we use to transition from $\Loss_2$ to $\Loss_1$. More formally:

\begin{equation*}
\Loss_{Huber}^{(M)}(r_i)=\begin{cases}
r_i^{2}, & \mbox{if $|r_i|\leq M$}\\
M (2|r_i|- M), & \mbox{otherwise}
\end{cases}
\end{equation*}

\item The loss function used in \textit{support vector regression (SVR)} is 
the $\epsilon$-insensitive loss, denoted by $\Loss_{\epsilon}^{(\epsilon)}$. Similar to 
Huber loss, there is a transition phase between small and large estimation errors. 
However, $\epsilon$-insensitive loss does not penalize the errors that are smaller 
than a threshold. Formally, we define $\epsilon$-insensitive loss as:

\begin{eqnarray*}
\Loss_{\epsilon}^{(\epsilon)}(r_i)&=& \bs{\max}(0,|r_i|-\epsilon)\\
&=&\begin{cases}
0, & \mbox{if $|r_i| \leq \epsilon$}\\
|r_i|-\epsilon, & \mbox{otherwise}
\end{cases}
\end{eqnarray*}
\end{itemize}

\begin{figure}
\centering
\includegraphics[width=\columnwidth]{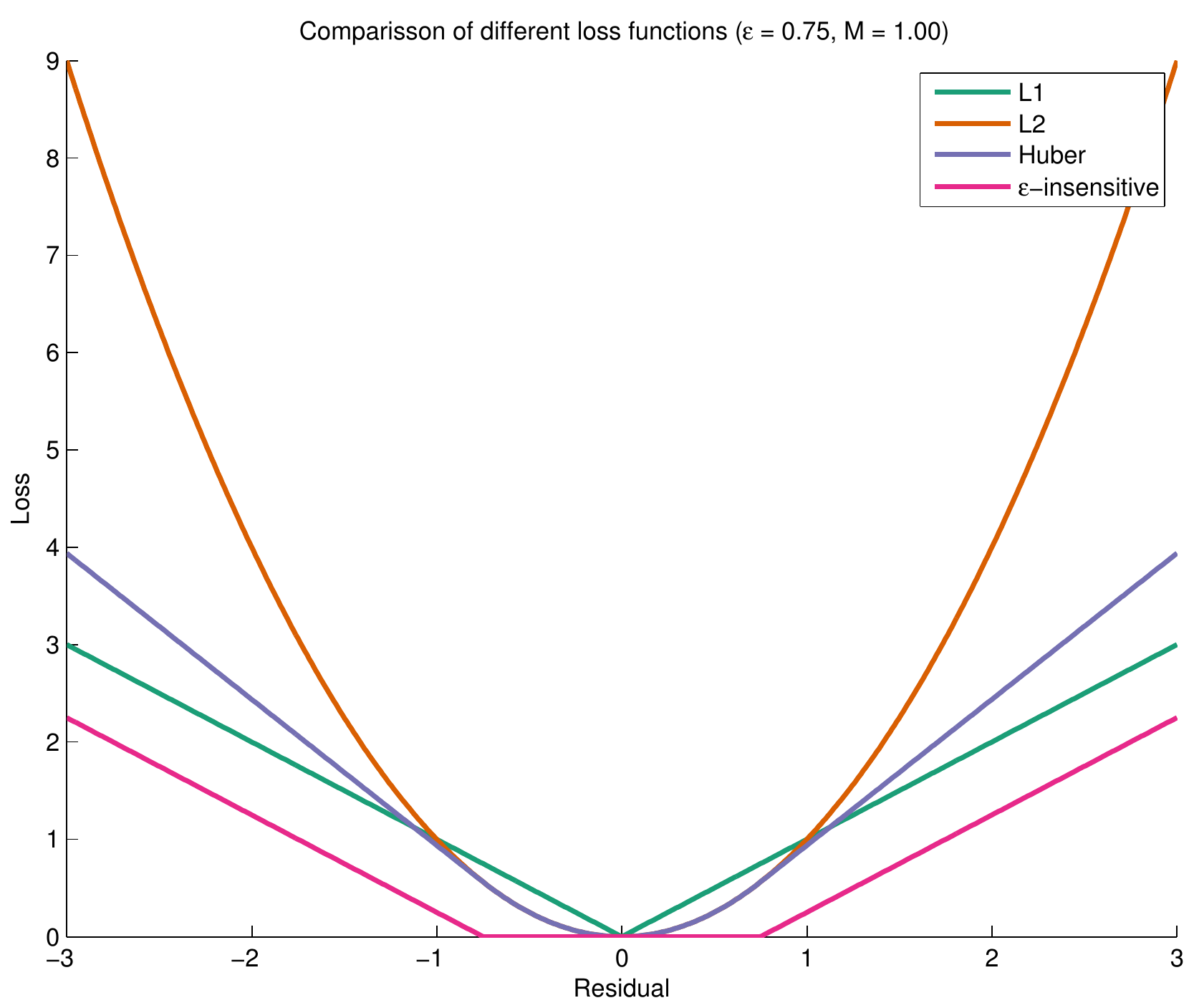}
\caption{Comparison of different loss functions}
\label{fig:loss_plot}
\end{figure}

Figure~\ref{fig:loss_plot} provides a visual representation of these loss functions, 
in which we use $M=1$ and $\epsilon = \frac{1}{2}$ for the Huber and 
$\epsilon$-insensitive loss functions, respectively. Note that for small residual 
values, $|r_i| \leq M = 1$, Huber and square loss are equivalent. However, outside
this region Huber loss becomes linear.

\subsubsection{Choice of Regularizers}
\label{sec:reg}

When the reference profile contains many cell-types that may not exist in mixtures,
or in cases where constitutive cell-types are highly correlated, regularizing the
objective function can sparsify the solution or enhance the conditioning of the problem.
We describe two commonly used regularizers here:

\begin{itemize}
\item The norm-2 regularizer is used to shrink the regression coefficient vector $\Vector{w}$
to ensure that it is as flat as possible. A common use of this regularizer is in
conjunction with $\Loss_2$ loss to remedy the multi-collinearity problem in classical
least squares regression. This regularizer is formally defined as:

\begin{equation}
\Reg_2(\Vector{w}) = \parallel\Vector{w}\parallel_2^2 = \sum_{i=1}^{k} w_i^2.
\end{equation}

\item Another common regularizer is the norm-1 regularizer, which is used to enforce
sparsity over $\Vector{w}$. Formally, it can be defined as:
\begin{equation}
\Reg_1(\Vector{w}) = \parallel\Vector{w}\parallel_1 = \sum_{i=1}^{k} |w_i|.
\end{equation}
\end{itemize}

In addition to these two regularizers, their combinations have also been 
introduced in the literature. One such example is \textit{elastic net}, which uses a 
convex combination of the two, that is $\Reg_{elastic}(\Vector{w}) = \alpha 
\Reg_1(\Vector{w}) + (1 - \alpha)\Reg_2(\Vector{w})$. Another example is \textit{group 
Lasso}, which, given a grouping $G$ among cell-types, enforces flatness among members of 
the group, while enhancing the sparsity pattern across groups. This regularizer function 
can be written as $\Reg_{group} = \sum_{G_i} \Loss_2( \Vector{w}(G_i))$, where 
$\Vector{w}(G_i)$ is the weight of cell-types in the $i^{th}$ group.

\subsection{Examples of objective functions used in practice}
\label{sec:obj_examples}

\subsubsection{Ordinary Least Squares (OLS)}

The formulation of OLS is based on squared loss, $\Loss_2$. Formally, we have:

\begin{eqnarray*}
\bs{\min}_{\Vector{w}} \{ \sum_{i=1}^m \Loss_2 (r_i) \} &=& \bs{\min}_{\Vector{w}} \{ \sum_{i=1}^m (y_{i}-\Vector{w}^{T}\Vector{x}_{i})^{2} \} \\ \nonumber
&=& \bs{\min}_{\Vector{w}}  \parallel y-\Matrix{X}\Vector{w}\parallel_{2}^{2}
\end{eqnarray*}

\noindent where row $i$ of the matrix $\Matrix{X}$, also known as the \textit{design 
matrix}, corresponds to $\Vector{x}_i$. This formulation has a closed form solution given 
by:

\begin{equation*}
\hat{\Vector{w}}=(\Matrix{X}^{T}\Matrix{X})^{-1}\Matrix{X}^{T}\Vector{y}
\end{equation*}

In this formulation, we can observe that norm-2 regularization is especially useful in 
cases where the matrix $\Matrix{X}$ is ill-conditioned and near-singular, that is, columns 
are dependent on each other. By shifting $\Matrix{X}^{T}\Matrix{X}$ towards the identity 
matrix, we ensure that the eigenvalues are farther from zero, which enhances the conditioning 
of the resulting combination.

\subsubsection{Ridge Regression}
\label{sec:Ridge}
One of the main issues with the OLS formulation is that the design matrix, $\Matrix{X}$, 
should have full column rank $k$. Otherwise, if we have highly correlated variables, 
the solution suffers from the \textit{multicollinearity} problem. This condition can be remedied by incorporating a norm-2 regularizer.
The resulting formulation, known as \textit{ridge regression}, is as follows:

\begin{alignat*}{2}
&\bs{\min}_{\Vector{w}} \{ \sum_{i=1}^m \Loss_2 (r_i)  + \lambda \Reg_2(\Vector{w})\}&\\
&= \bs{\min}_{\Vector{w}} \parallel y-\Matrix{X}\Vector{w}\parallel_{2}^{2} + \lambda\parallel \Vector{w}\parallel_{2}^2&
\end{alignat*}

Similar to OLS, we can differentiate w.r.t. $\Vector{w}$ to find the close form solution
for Ridge regression given by:

\begin{equation*}
\hat{\Vector{w}} = (\Matrix{X}^T\Matrix{X} + \lambda\Matrix{I})^{-1}\Matrix{X}^T\Vector{y}
\end{equation*}

\subsubsection{Least Absolute Selection and Shrinkage Operator (LASSO) Regression}
\label{sec:LASSO}

Combining the OLS with a norm-1 regularizer, we have the LASSO formulation:

\begin{alignat*}{2}
&\bs{\min}_{\Vector{w}} \{ \sum_{i=1}^m \Loss_2 (r_i)  + \lambda \Reg_1(\Vector{w})\}&\\
&= \bs{\min}_{\Vector{w}} \parallel y-\Matrix{X}\Vector{w}\parallel_{2}^{2} + \lambda \parallel \Vector{w}\parallel_{1}&
\end{alignat*}

This formulation is especially useful for producing sparse solutions by introducing
zero elements in vector $\Vector{w}$.  However, while being convex, it does not have a
closed form solution.

\subsubsection{Robust Regression}

It is known that $\Loss_2(\Vector{r})$ is dominated by the largest 
elements of the residual vector $\Vector{r}$, which makes it sensitive to outliers. To 
remedy this problem, different robust regression formulations have been proposed that use 
alternative loss functions. Two of the best-known formulations are based on the 
$\Loss_1$ and $\Loss_{huber}$ loss functions. The $\Loss_1$ formulation can be written 
as:

\begin{eqnarray*}
\bs{\min}_{\Vector{w}} \{  \sum_{i=1}^m \Loss_1 (r_i) \} &=& \bs{\min}_{\Vector{w}} \{ \sum_{i=1}^m |y_{i}-\Vector{w}^{T}\Vector{x}_{i}| \} \\ \nonumber
&=& \bs{\min}_{\Vector{w}} \parallel y-\Matrix{X}\Vector{w}\parallel_{1}
\end{eqnarray*}

However, for the Huber loss function, while it can be defined similarly, it is usually 
formulated as an alternative convex Quadratic Program (QP):

\begin{alignat}{2}
\bs{\min}_{\Vector{x}, \Vector{z}, \Vector{t}} & \{\frac{1}{2} \parallel \Vector{z}\parallel_{2}^{2} +  M \Vector{1}^T\Vector{t} \}\nonumber\\
\mbox{Subject to:} & - \Vector{t} \leq \Matrix{X}\Vector{w} - \Vector{y} - \Vector{z}  \leq \Vector{t}
\end{alignat}

\noindent which can be solved more efficiently using the following equivalent QP 
variant \cite{Mangasarian2000}:

\begin{alignat}{2}
\bs{\min}_{\Vector{x}, \Vector{z}, \Vector{r}, \Vector{s}} & \{\frac{1}{2} \parallel \Vector{z}\parallel_{2}^{2} +  M\Vector{1}^T(\Vector{r} + \Vector{s}) \}\nonumber\\
\mbox{Subject to:} & \begin{cases}
 \Matrix{X}\Vector{w} - \Vector{y} - \Vector{z}  = \Vector{r} - \Vector{s}&\\
0 \leq \Vector{r}, \Vector{s}
\end{cases}
\end{alignat}

In both of these formulations, the scalar $M$ corresponds to half-length parameter of the 
Huber's loss function.

\subsubsection{Support Vector Regression}

In machine learning, Support Vector Regression (SVR) is a commonly used technique
that aims to find a regression by maximizing the margins around the estimated separator
hyperplane from the closest data points on each side of it. This margin provides the
region in which estimation errors are ignored. SVR has been recently used to deconvolve
biological mixtures, where it has been shown to outperform other methods \cite{Newman2015}.
One of the variants of SVR is $\epsilon$-SVR, in which parameter $\epsilon$ defines the
margin, or the $\epsilon$-tube. The primal formulation of $\epsilon$-SVR with linear
kernel can be written as \cite{Vapnik1998}:

\begin{alignat}{2}
\bs{\min}_{\Vector{w}, \Vector{\xi}^{+}_{i}, \Vector{\xi}^{-}_{i}} & \{\frac{1}{2} \parallel \Vector{w}\parallel_{2}^{2} +  C \sum_{i=1}^{m}(\Vector{\xi}_{i}^{+}+\Vector{\xi}^{-}_{i}) \}\nonumber\\
\mbox{Subject to:} & \begin{cases}
y_{i} - \Vector{w} \cdot\Vector{x}_{i}\leq\epsilon+\Vector{\xi}_{i}^{+}&\\
-(\epsilon+\Vector{\xi}^{-}_{i}) \leq y_{i} - \Vector{w}\cdot\Vector{x}_{i}&\\
0 \leq \Vector{\xi}_{i}^{+}, \Vector{\xi}^{-}_{i}&   
\end{cases}
\label{eq:SVR_primal}
\end{alignat}

\noindent in which, given the \textit{unit norm assumption} introduced in 
Section~\ref{sec:decon_def}, we assume that $b =0$. The dual problem for the primal in 
Equation~\ref{eq:SVR_primal} can be written in matrix form as:

\begin{IEEEeqnarray}{lll}
\bs{\max}_{\Vector{\alpha}^{+}, \Vector{\alpha}^{-} } & \Big\{\Vector{1}^T\big((\Vector{\alpha}^{+} - \Vector{\alpha}^{-}) \odot \Vector{y}\big) &\nonumber\\
& -\epsilon\Vector{1}^T(\Vector{\alpha}^{+} + \Vector{\alpha}^{-})& \nonumber\\
& - (\Vector{\alpha}^{+} - \Vector{\alpha}^{-})^T\Matrix{K}(\Vector{\alpha}^{+} - \Vector{\alpha}^{-}) \Big\}&\nonumber\\
&\mbox{Subject to:}  \begin{cases}
\Vector{1}^T(\Vector{\alpha}^{+} - \Vector{\alpha}^{-}) = 0& \\
0 \leq\Vector{\alpha}^{+},  \Vector{\alpha}^{-} \leq C
\end{cases}&
\label{eq:dont_use_multline}
\end{IEEEeqnarray}

In this formulation, $\Vector{1}$ is a vector of all ones, $\odot$ is the element-wise 
product, and $\Matrix{K}$ is the kernel matrix defined as $\Matrix{K} = 
\Matrix{X}\Matrix{X}^T$. The dual formulation is often used to solve 
$\epsilon$-SVR, because it can be easily extended to use different kernel functions to 
map $\Vector{x}_i$ to a $d$-dimensional non-linear feature space. Additionally, when $m 
\ll k$, such as the case of high-dimensional feature spaces, it provides a better way to 
solve the SVR problem. However, the primal problem provides a more straightforward interpretation.
In addition, in the case where $k \ll m$, it provides superior performance. To show the
similarity with Equation~\ref{eq:linReg}, we can rewrite Equation~\ref{eq:SVR_primal} 
using the $\epsilon$-insensitive loss function as follows:

\begin{equation}
\bs{\min}_{\Vector{w}} \{ \sum_{i = 1}^m \mathcal{L}_{\epsilon}(y_{i}-\Vector{w}^T\Vector{x}_{i})  +  \lambda \Reg_2(\Vector{w})  \}
\label{eq:SVR_primal_alt}
\end{equation}

\noindent where $\lambda = \frac{1}{2C}$ \cite{Smola2004}.


\subsection{Overview of Prior in silico Deconvolution Methods}
\label{sec:overview}

A majority of existing deconvolution methods fall into two groups -- they either 
use a regression-based framework to compute $\Matrix{G}$, $\Matrix{C}$, or both; or
perform statistical inference over a  probabilistic model. Abbas \textit{et al.} \cite{Abbas2009}
present one of the early regression-based methods for estimating $\Matrix{C}$. This method
is designed to identify cell-type concentrations  from
a known reference profile of immune cells. Their method is based on Ordinary 
Least Squares (OLS) regression and does not consider either non-negativity or 
sum-to-one constraints explicitly, but rather it enforces these constraints implicitly
after the optimization procedure.
An extension of this approach is proposed by Qiao \textit{et al.} \cite{Qiao2012}, which 
uses non-negative least squares (NNLS) to explicitly enforce non-negativity as part of 
the optimization. Gong \textit{et al.} \cite{Gong2011} present a quadratic programming 
(QP) framework to explicitly encode both constraints in the optimization problem 
formulation. They also propose an extension to this method, called DeconRNASeq, 
which applies the same QP framework to RNASeq datasets. More recently Newman 
\textit{et al.} \cite{Newman2015} propose robust linear regression (RLR) and 
$\nu$-SVR regression instead of $\Loss_2$ based regression, which is highly 
susceptible to noise. Digital cell quantification (DCQ) \cite{Altboum2014} is another 
approach designed for monitoring the immune system during infection.
Compared to prior methods, DCQ forces sparsity by combining 
$\Reg_2$ and $\Reg_1$ regularization into an \textit{elastic net}. This 
regularization is essential for successfully identifying the subset of 
active cells at each stage, given the larger number of cell-types included in their 
panel (213 immune cell sub-populations). In contrast to these techniques, Shen-Orr 
\textit{et al.} \cite{Shen-Orr2010} propose a method, call \textit{csSAM}, which is 
specifically designed to identify genes that are differentially expressed among
purified cell-types. The core of this method is regression over matrix $\Matrix{C}$
to estimate matrix $\Matrix{G}$.

Full regression-based methods use variations of block-coordinate descent 
to successively identify better estimates for both $\Matrix{C}$ and $\Matrix{G}$ \cite{Kim2013}.
Venet \textit{et al.} \cite{Venet2001} present one of the early methods in this class, which
uses an NMF-like method coupled with a heuristic to decorrelate columns of $\Matrix{G}$ in
each iteration. Repsilber \textit{et al.} \cite{Repsilber2010} propose an algorithm called 
\textit{deconf}, which uses alternating non-negative least squares (ANLS) for solving 
NMF, without the decorrelation step of Vennet \textit{et al.}, while implicitly applying 
constraints on $\Matrix{C}$ and $\Matrix{G}$ at each iteration. Inspired by the work of Pauca 
\textit{et al.} on hyperspectral image deconvolution \cite{Pauca2006}, Zuckerman 
\textit{et al.} \cite{Zuckerman2013} propose an NMF method based on the Frobenius 
norm for gene expression deconvolution. They use gradient descent to solve for $\Matrix{C}$
and $\Matrix{G}$ at each step, which 
converges to a local optimum of the objective function. Given that the expression domain 
of cell-type specific markers is restricted to unique cells in the reference profile, 
Gaujoux \textit{et al.} \cite{Gaujoux2012} present a semi-supervised NMF (ssNMF) method 
that explicitly enforces an orthogonality constraint at each iteration over the subset 
of markers in the reference profile. This constraint both enhances the convergence of the 
NMF algorithm, and simplifies the matching of columns in the estimated cell-type 
expression to the columns of the reference panel, $\Matrix{G}$. The Digital Sorting Algorithm 
(DSA) \cite{Zhong2013} works as follows: if 
concentration matrix $\Matrix{C}$ is known a priori, it directly uses quadratic 
programming (QP) with added constraints on the lower/upper bound of gene expressions
to estimate matrix $\Matrix{G}$. Otherwise, if fractions are also unknown, it
uses the average expression of given marker genes that are only expressed in one cell-type,
combined with the \textit{STO} constraint, to estimate concentrations matrix $\Matrix{C}$
first. Population-specific expression analysis (PSEA) \cite{Kuhn2011} performs a linear
least squares regression to estimate quantitative measures of cell-type-specific
expression levels, in a similar fashion as the update equation for estimating
$\hat{\Matrix{G}}$ in Equation~\ref{eq:ANLS}. In cases where the matrix $\Matrix{C}$ is
not known a priori, \textit{PSEA} exploits the average expression of marker genes
that are exclusively expressed in one of the reference profiles as \textit{reference
signals} to track the variation of cell-type fractions across multiple mixture samples.

In addition to regression-based methods, a large class of methods is based on 
probabilistic modeling of gene expression. Erikkila \textit{et al.} \cite{Erkkila2010} 
introduce a method, called \textit{DSection}, which formulates the deconvolution problem
using a Bayesian model. It incorporates a Bayesian prior over the noisy observation of given
concentration parameters to account for their uncertainty, and employs a MCMC sampling
scheme to estimate the posterior distribution of the parameters/latent variables,
including $\Matrix{G}$ and a denoised version of $\Matrix{C}$. The in-silico NanoDissection
method \cite{Ju2013} uses a classification algorithm based on linear
SVM coupled with an iterative adjustment process to refine a set of provided, positive
and negative, marker genes and infer a ranked list of genome-scale predictions for
cell-type-specific markers. Quon \textit{et al.} \cite{Quon2013} propose a probabilistic
deconvolution method, called \textit{PERT}, which estimates a global, multiplicative
perturbation vector to correct for the differences between provided reference profiles
and the true cell-types in the mixture. \textit{PERT} formulates the deconvolution
problem in a similar framework as Latent Dirichlet Allocation (LDA), and uses
the conjugate gradient descent method to cyclically optimize the joint likelihood
function with respect to each latent variable/parameter. Finally, microarray microdissection
with analysis of differences (MMAD) \cite{Liebner2014} incorporates the concept of
the \textit{effective RNA fraction} to account for source and sample-specific bias in
the cell-type fractions for each gene. They propose different strategies depending
on the availability of additional data sources. 
In cases where no additional information is available, they identify genes with the
highest variation in mixtures as markers and assign them to different reference
cell-types using k-means clustering, and finally use these \textit{de novo} markers
to compute cell-type fractions. \textit{MMAD} uses a MLE approach over the residual
sum of squares to estimate unknown parameters in their formulation.

\section{Results and Discussion}
\label{sec:results}

We now present a comprehensive evaluation of various formulations for solving deconvolution
problems. Some of these algorithimic combinations have been proposed in literature, while
others represent new algorithmic choices. We systematically assess the impact of
these algorithmic choices on the performance of in-silico deconvolution.

\subsection{Datasets}
\label{sec:datasets}

\begin{enumerate}

\item \textbf{In vivo mixtures with known percentages}: We use a total of five datasets with 
known mixtures. We use CellMix to download and normalize these datasets \cite{Gaujoux2013},
which uses the \textit{soft} format data available from Gene Expression Omnibus (GEO).

\begin{itemize}

\item \textbf{BreatBlood} \cite{Gong2011} (GEO ID: \textit{GSE29830}): Breast and blood from human
specimens are mixed in three different proportions and each of the mixtures is measured three times, 
with a total of nine samples.

\item \textbf{CellLines} \cite{Abbas2009} (GEO ID: \textit{GSE11058}): Mixture of human cell 
lines Jurkat (T cell leukemia), THP-1 (acute monocytic leukemia), IM-9 (B lymphoblastoid 
multiple myeloma) and Raji (Burkitt B-cell lymphoma) in four different concentrations, each of 
which is repeated three times, resulting in a total of 12 samples.

\item \textbf{LiverBrainLung} \cite{Shen-Orr2010} (GEO ID: \textit{GSE19830}): This dataset 
contains three different rat tissues, namely brain, liver, and lung tissues, which are mixed in 
11 different concentrations with each mixture having three technical replicates, for a total of 
33 samples.


\item \textbf{RatBrain} \cite{Kuhn2011} (GEO ID: \textit{GSE19380}):  This contains four different 
cell-types, namely rat's neuronal, astrocytic, oligodendrocytic and microglial cultures, and two
replicates of five different mixing proportions, for a total of 10 samples.

\item \textbf{Retina} \cite{Siegert2012} (GEO ID: \textit{GSE33076}):  This dataset pools
together retinas from two different mouse lines and mixed them in eight different
combinations and three replicates for each mixture, resulting in a total of 24 samples.

\end{itemize}

\item \textbf{Mixtures with available cell-sorting data through flow-cytometry}: For this 
experiment, we use two datasets available from Qiao \textit{et al.} \cite{Qiao2012}. We 
directly download these datasets from the supplementary material of the paper. These datasets
are post-processed by the supervised normalization of microarrays (SNM) method to correct for
batch effects. Raw expression profiles are also available for download under GEO ID \textit{GSE40830}.
This dataset contains two sub-datasets:

\begin{itemize}

\item \textbf{PERT\_Uncultured}: This dataset contains uncultured human cord blood 
mono-nucleated and lineage-depleted (Lin-) cells on the first day.

\item \textbf{PERT\_Cultured}: This dataset contains culture-derived lineage-depleted
human blood cells after four days of cultivation.

\end{itemize}
\end{enumerate}

Table~\ref{table:ds_stats} summarizes overall statistics related to each of these datasets.

\begin{table}[!h]
\caption{Summary statistics of each dataset}
\centering
\begin{tabular}{l|c|c|c}
Dataset	&\# features	&\# samples	&\# references\\ \hline
BreastBlood	&54675	&9	&2\\
CellLines	&54675	&12	&4\\
LiverBrainLung	&31099	&33	&3\\
PERT\_Cultured	&22215	&2	&11\\
PERT\_Uncultured	&22215	&4	&11\\
RatBrain	&31099	&10	&4\\
Retina	&22347	&24	&2\\
\end{tabular}
\label{table:ds_stats}
\end{table}

\subsection{Evaluation Measures}
\label{sec:eval_measures}

Let us denote the actual and estimated coefficient matrices by $\Matrix{C}$ and 
$\hat{\Matrix{C}}$ , respectively. We first normalize these measures to ensure each column 
sums to one. Then, we define the corresponding percentages as $\Matrix{P} = 100 \times 
\Matrix{C}_{norm}$ and $\hat{\Matrix{P}} = 100 \times \hat{\Matrix{C}}_{norm}$. Finally, let
 $r_{jk} = p_{jk} - \hat{p}_{jk}$ be the residual estimation error of cell-type $k$ in sample $j$.
Using this notation, we can define three commonly used measures of estimation error
as follows:

\begin{enumerate}

\item \textbf{Mean absolute difference (mAD)}: This is among the easiest measures to 
interpret.  It is defined as the average of all differences for different cell-type 
percentages in different mixture samples. More specifically:

\begin{equation*}
mAD = \frac{1}{p \times q} \sum_{j = 1}^{p} \sum_{k = 1}^{q} |r_{jk} |
\end{equation*}

\item \textbf{Root mean squared distance (RMSD)}: This measure is one of the most commonly used
distance functions in the literature. It is formally defined as:

\begin{equation*}
mAD = \sqrt{\frac{1}{p \times q} \sum_{j = 1}^{p} \sum_{k = 1}^{q} r_{jk}^2}
\end{equation*}

\item \textbf{Pearson's correlation distance}: Pearson's correlation measures the linear 
dependence between estimated and actual percentages. Let us vectorize percentage matrices as 
$\Vector{p} = \textbf{vec}(\Matrix{P})$ and $\hat{\Vector{p}} = 
\textbf{vec}(\hat{\Matrix{P}})$. Using this notation, the correlation between these two 
vectors is defined as:

\begin{equation}
\rho_{\Vector{p}, \hat{\Vector{p}}} = \frac{\textbf{cov}(\Vector{p}, \hat{\Vector{p}})}{\sigma(\Vector{p})\sigma(\hat{\Vector{p}})}
\end{equation}

\noindent where $\textbf{cov}$ and $\sigma$ correspond to covariance and standard variation of 
vectors, respectively. Finally, we define the correlation distance measure as $R^2D = 1 - 
\rho_{\Vector{p}, \hat{\Vector{p}}}$.

\end{enumerate}

\subsection{Implementation}

All codes and experiments have been implemented in Matlab. To implement different formulations of 
the deconvolution problem, we used CVX, a package for specifying and solving convex 
programs \cite{gb08, cvx}. We used Mosek together with CVX, which is a high-performance solver 
for large-scale linear and quadratic programs \cite{mosek}. All codes and datasets are freely 
available at \url{github.com/shmohammadi86/DeconvolutionReview}.

\subsection{Effect of Loss Function and Constraint Enforcement on Deconvolution Performance}
\label{sec:loss_effect}

We perform a systematic evaluation of the four different loss functions introduced in 
Section~\ref{sec:loss}, as well as implicit and explicit enforcement of \textit{non-negativity 
(NN)} and \textit{sum-to-one (STO)} constraints over the concentration matrix 
($\Matrix{\hat{C}}$), on the overall performance of deconvolution methods for each dataset. 
There are 16 configurations of loss functions/constraints for each test case. 
Additionally, for Huber and Hinge loss functions, where $M$ and $\epsilon$ are unknown, 
we perform a grid search with 15 values in multiples of 10 spanning the range $\{ 10^{-7}, \cdots, 10^{7} \}$ 
to find the best values for these parameters. In order to evaluate an upper bound on the 
``potential" performance of these two loss functions, we use the true concentrations in each 
sample, $\Matrix{c}$, to evaluate each parameter choice. In practical applications, 
the RMSD of residual error between $\Vector{m}$ and $\Matrix{G}\Vector{\hat{c}}$ is often
used to select the optimal parameter. This is not always in agreement with the choice made 
based on known $\Vector{c}$.

\begin{figure}
\centering
\includegraphics[width=.9\columnwidth]{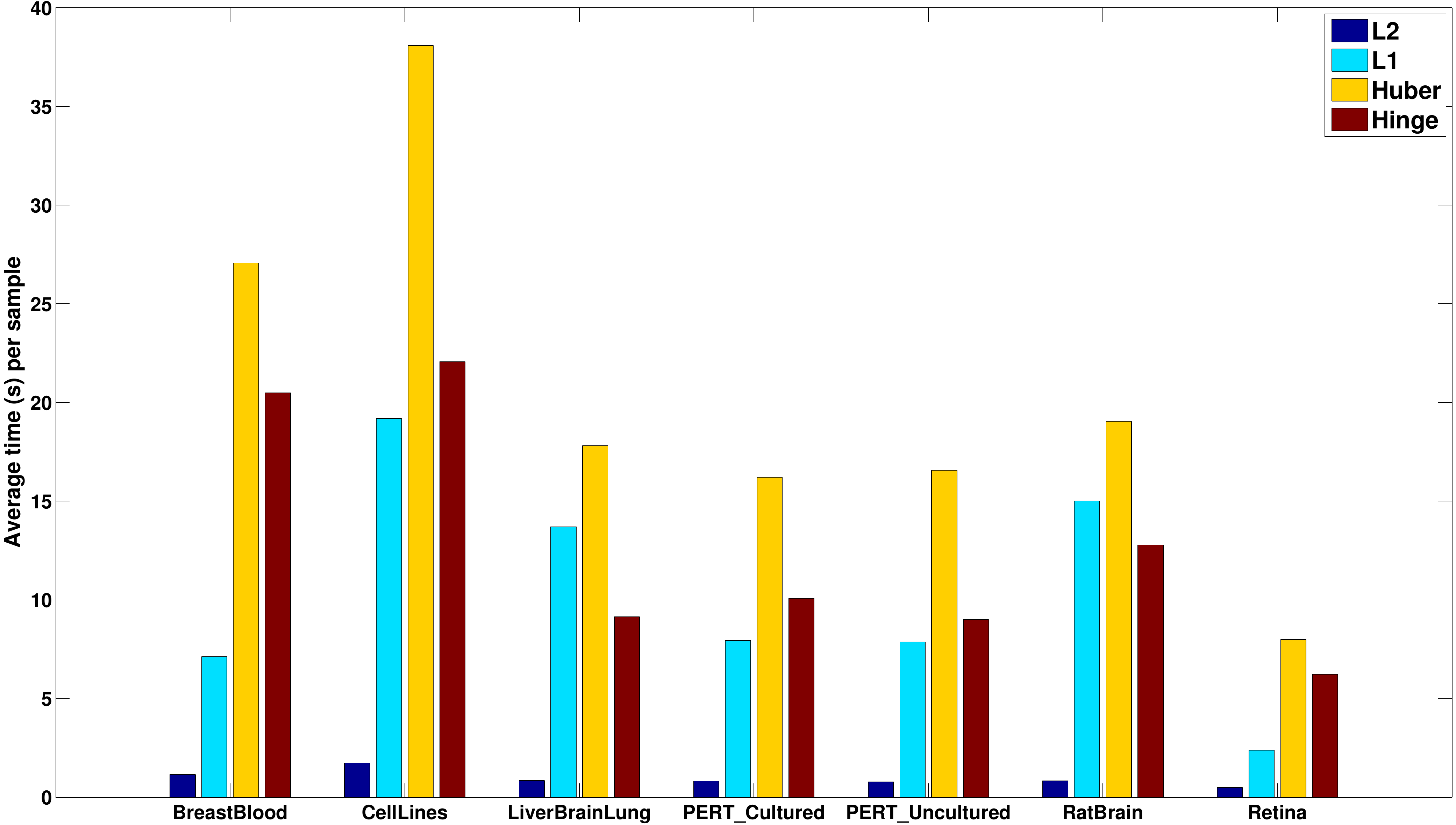}
\caption{Average computational time for each loss function in different datasets}
\label{exp1_timing}
\end{figure}

For each test dataset, we compute the three evaluation measures defined in 
Section~\ref{sec:eval_measures}. Additionally, for each of these measures, we compute an 
empirical \emph{p}-value by sampling random concentrations from a Uniform distribution and 
enforcing \textit{NN} and \textit{STO} constraints on the resulting random sample. In our 
study, we sampled $10,000$ concentrations for each dataset/measure, which results in a lower 
bound of $10^{-4}$ on the estimated \emph{p}-values. Figure~\ref{exp1_timing} presents the 
time each loss function takes to compute per sample, averaged over all constraint combinations. 
The actual times taken for Huber and Hinge losses are roughly 15 times those reported here, 
which is the number of experiments performed to find the optimal parameters for these loss functions.
From these results, $\Loss_2$ can be observed to have the fastest computation time,
whereas $\Loss_{Huber}$ is the slowest.
Measures $\Loss_1$ and $\Loss_{Hinge}$ fit in between these two extremes, with $\Loss_{1}$
being faster the majority of times. We can directly compare these computation times,
because we formulate all methods within the same framework; thus, differences in
implementations do not impact direct comparisons.

\begin{figure}[!t]
\centering
\includegraphics[width=0.8\columnwidth]{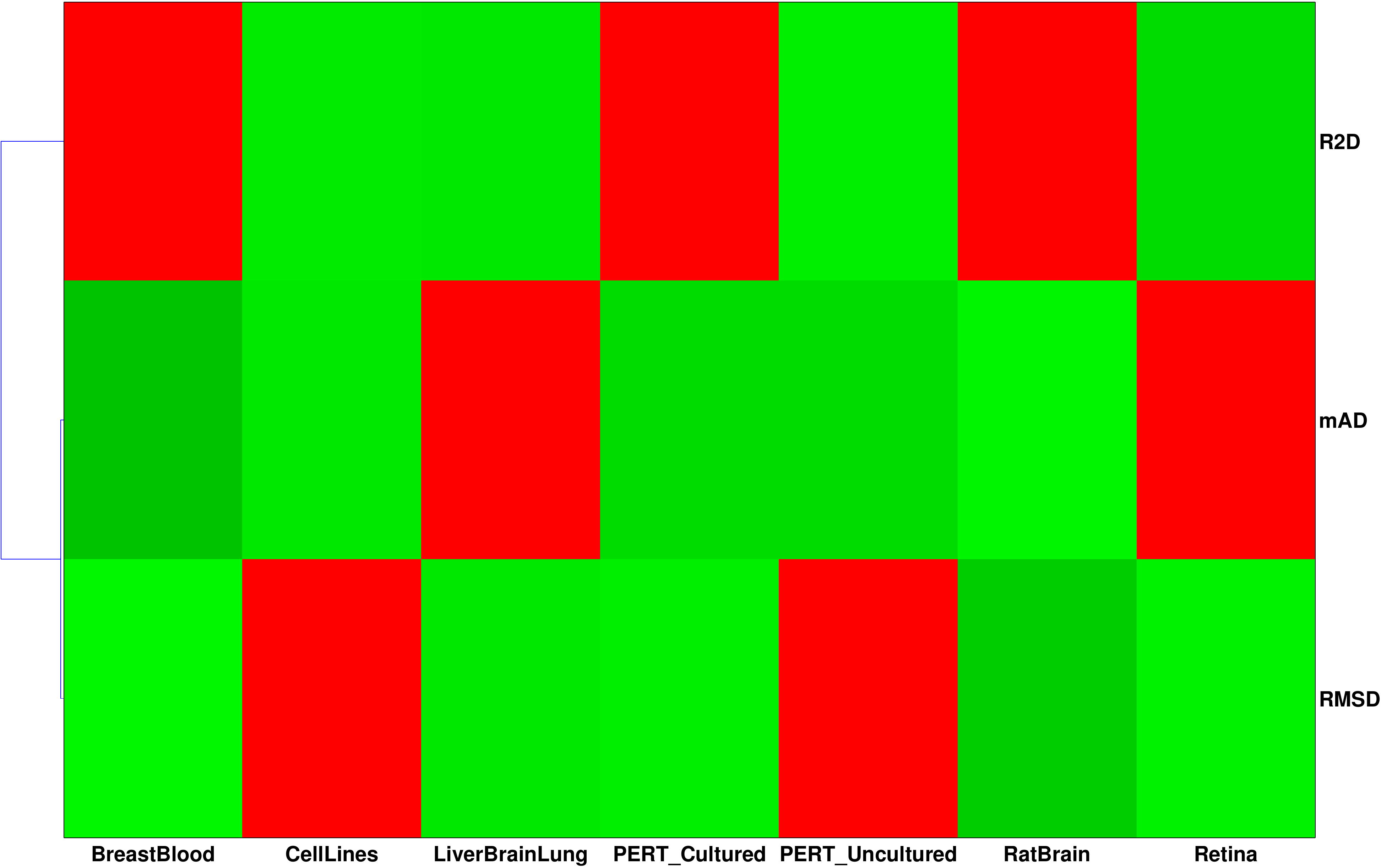}
\caption{Agreement among different evaluation measures across different datasets}
\label{fig:measure_corr}
\end{figure}

Computation time, while important, is not the critical measure in our evaluation. The 
true performance of a configuration (selection of loss function and constraints) is measured
by its estimation error. In order to rank different configurations, we first assess the agreement
among different measures. To this end, we evaluate each dataset as follows: for each experiment, we 
compute $\mathbf{mAD, RMSD}$, and $\mathbf{R^2D}$ independently. Then, we use \textit{Kendall} 
rank correlation, a non-parametric hypothesis test for statistical dependence between two 
random variables, between each pair of measures and compute a log-transformed \emph{p}-value 
for each correlation. Figure~\ref{fig:measure_corr} shows the agreement among these 
measures across different datasets. Overall, $RMSD$ and $mAD$ measures show higher 
consistency, compared to $R^2D$ measure. However, the $mAD$ measure is easier to interpret as a 
measure of percentage loss for each configuration. Consequently, we choose this measure for our 
evaluation in this study.

\begin{figure}
\centering
\includegraphics[width=.9\columnwidth]{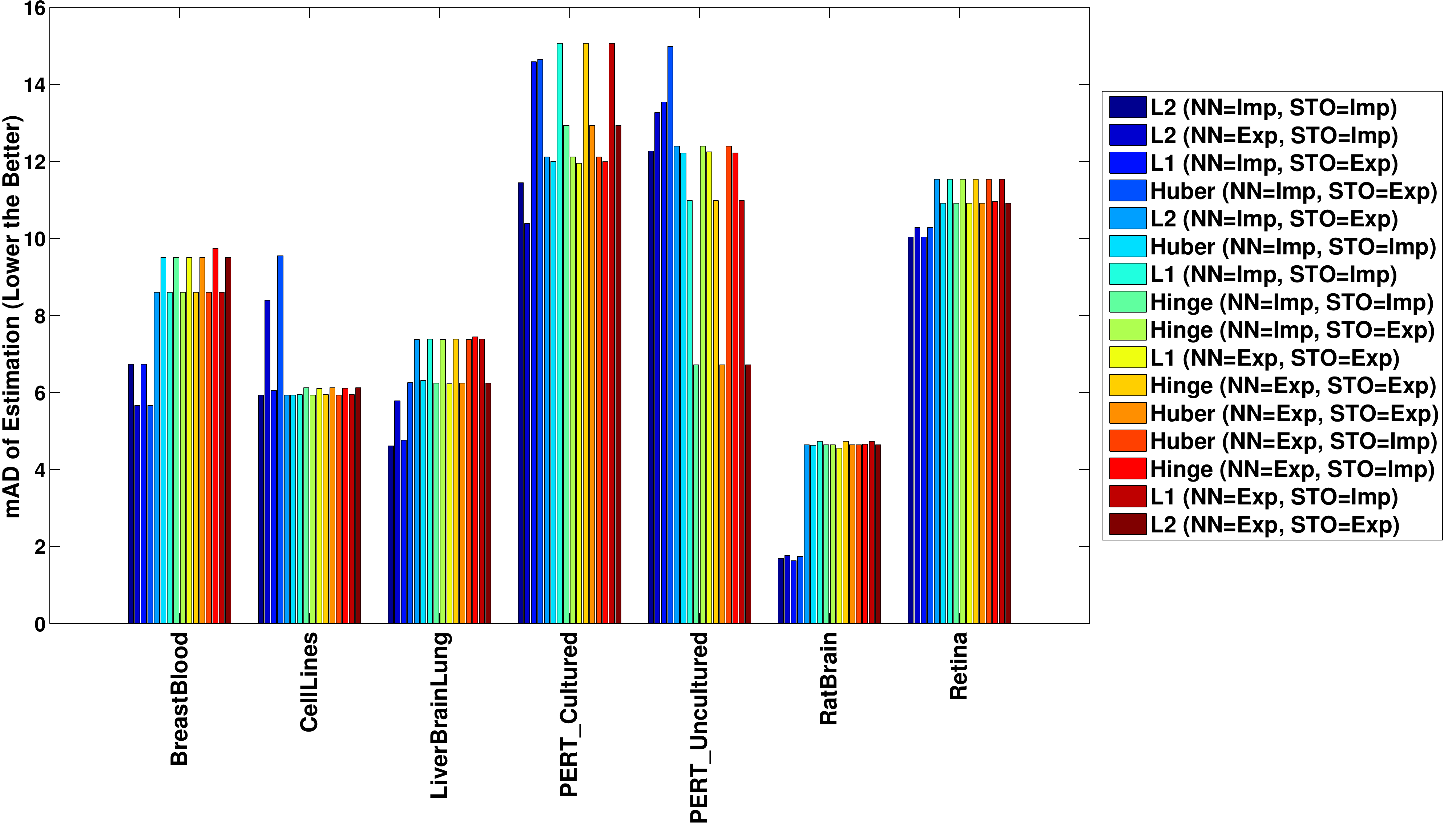}
\caption{Overall performance of different loss/constraints combinations over all datasets}
\label{exp1_mAD}
\end{figure}

Using $mAD$ as the measure of performance, we evaluate each configuration over each dataset 
and sort the results. Figure~\ref{exp1_mAD} shows various
combinations for each dataset. The \textbf{RatBrain}, \textbf{LiverBrainLung}, 
\textbf{BreastBlood}, and \textbf{CellLines} datasets achieve high performance. Among these 
datasets, \textbf{RatBrain}, \textbf{LiverBrainLung}, and \textbf{BreastBlood} had the $\Loss_2$ 
loss function as the best configuration, with the \textbf{CellLines} dataset being less 
sensitive to the choice of the loss function. Another surprising observation is that for the 
majority of configurations, enforcing the \textit{sum-to-one} constraint worsens the results. 
We investigate this issue in greater depth in Section~\ref{sec:STO_effect}.

\begin{figure}[!t]
\centering
\includegraphics[width=0.8\columnwidth]{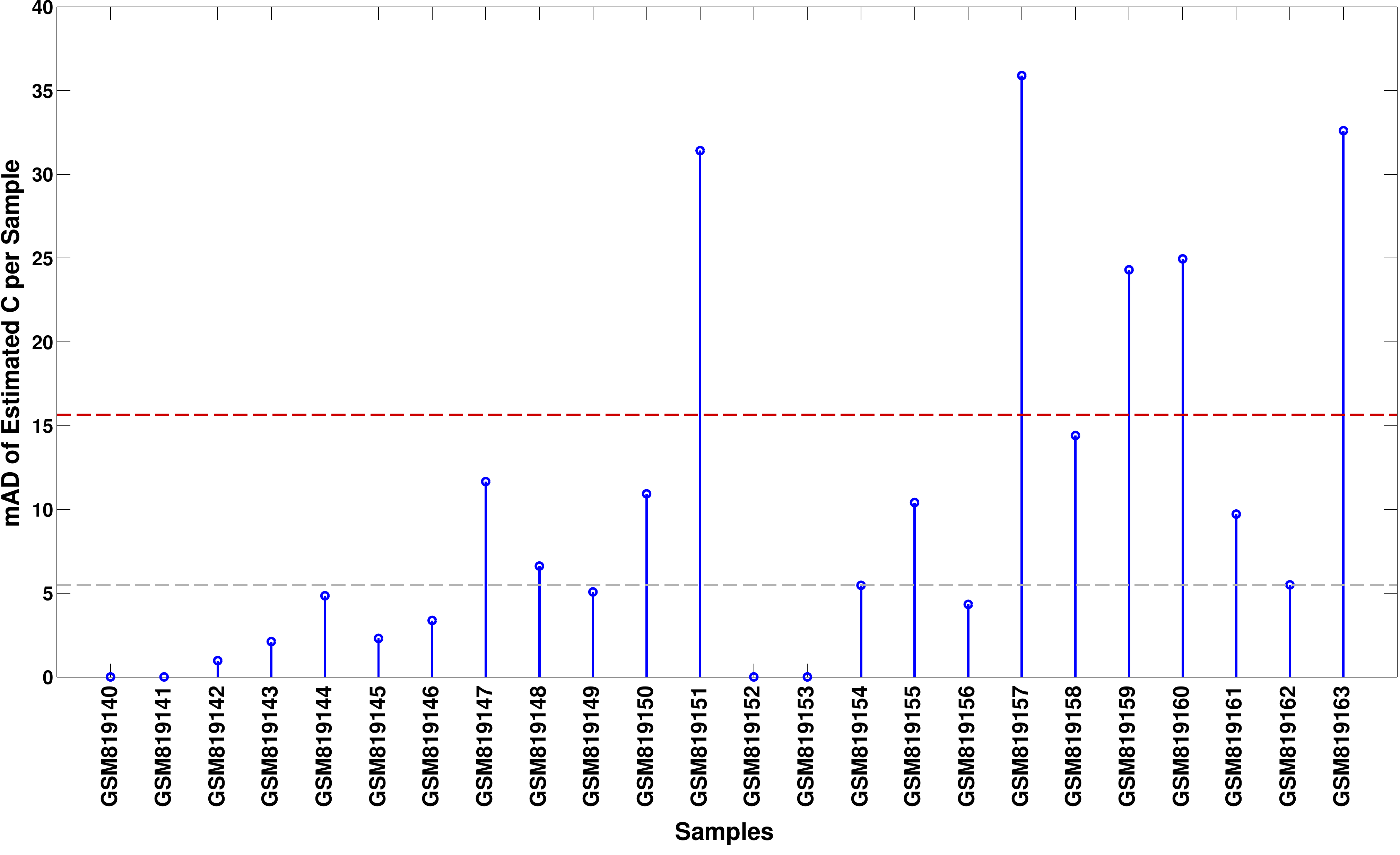}
\caption{Sample-based error of the Retina dataset, based on $\Loss_2$ with explicit \textit{NN} and \textit{STO}} 
\label{fig:Retina_QC}
\end{figure}

For \textbf{Retina}, as well as both \textbf{PERT} datasets, the overall performance is worse 
than the other datasets. In the case of \textbf{PERT}, this is expected, since the flow-sorted 
proportions are used as an estimate of cell-type proportions. Furthermore, the reference profiles
come from a different study and therefore have greater difference with the true cell-types in the 
mixture. However, the \textbf{Retina} dataset exhibits unusually low performance, which may be
attributed to multiple factors. As an initial investigation, we performed a quality control
(QC) over different samples to see if errors are similarly distributed across samples. 
Figure~\ref{fig:Retina_QC} presents per-sample error, measured by $mAD$, with median and 
median absolute deviation (MAD) marked accordingly. Interestingly, for the $4^{th}, 6^{th}$, and $8^{th}$
mixtures, the third replicate has much higher error than the rest. In the expression matrix, we observed 
a lower correlation between these replicates and the other two replicates in the batch. 
Additionally, for the $7^{th}$ mixture, all three replicates show high error rates.
We expand on these results in later sections to identify additional 
reasons that contribute to the low deconvolution performance of the
\textbf{Retina} dataset.

Finally, we note that in all test cases the performance of $\Loss_1, \Loss_{Huber}$, and 
$\Loss_{Hinge}$ are comparable, while $\Loss_{Huber}$ and $\Loss_{Hinge}$ needed an 
additional step of parameter tuning. Consequently, we only consider $\Loss_1$ as a representative 
of this ``robust'' group of loss functions in the rest of our study.

\subsection{Agreement of Gene Expressions with Sum-to-One (STO) Constraint}
\label{sec:STO_effect}

Considering the lower performance of configurations that explicitly enforce STO constraints, we aim 
to investigate whether features (genes) in each dataset respect this constraint. Under the
\textit{STO} and \textit{NN} constraints, we use simple bounds for identifying violating
features, for which there is no combination of concentration values that can satisfy both
\textit{STO} and \textit{NN}. Let $\Vector{m}(i)$ be the expression value of the 
$i^{th}$ gene in the given mixture, and $\Matrix{G}(i, 1), \cdots, \Matrix{G}(i, q)$ be the 
corresponding expressions in different reference cell-types. Let $\Matrix{G}_{min}(i) = 
\textbf{min} \{ \Matrix{G}(i, 1), \cdots, \Matrix{G}(i, q) \}$ and $\Matrix{G}_{max}(i) = 
\textbf{max} \{ \Matrix{G}(i, 1), \cdots, \Matrix{G}(i, q) \}$. Given that all concentrations 
are bound between $0 \leq \Vector{c}(k) \leq 1; \forall 1 \leq k \leq k$, the minimum and 
maximum values that an estimated mixture value for the $i^{th}$ gene can attain are 
$\Matrix{G}_{min}(i)$ and $\Matrix{G}_{max}(i) $, respectively (by setting $\Vector{c}(k)=1$ 
for min/max value, and $0$ everywhere else). Using this notation, we can identify features 
that violate \textit{STO} as follows:

\begin{alignat*}{3}
& \Vector{m}(i) \leq \Matrix{G}_{min}(i)   \quad && \forall 1 \leq i \leq n           \quad && \text{ \{Violating reference\}}\\
& \Matrix{G}_{max}(i) \leq \Vector{m}(i)    \quad && \forall 1 \leq i \leq n     \quad && \text{ \{Violating mixture\} }\\
\end{alignat*}

The first condition holds because expression values in reference profiles are so large
that we need the sum of concentrations to be lower than one to be able to match the
corresponding gene expression in the mixture. The second condition holds in cases where
the expression of a gene in the mixture is so high that we need the sum of concentrations
to be greater than one to be able to match it. In other words, for feature $i$, these
constraints identify extreme expression values in reference 
profiles and mixture samples, respectively. Using these conditions, we compute the total 
number of features violating \textit{STO} condition in each dataset. 

\begin{figure}[!t]
\centering
\includegraphics[width=0.8\columnwidth]{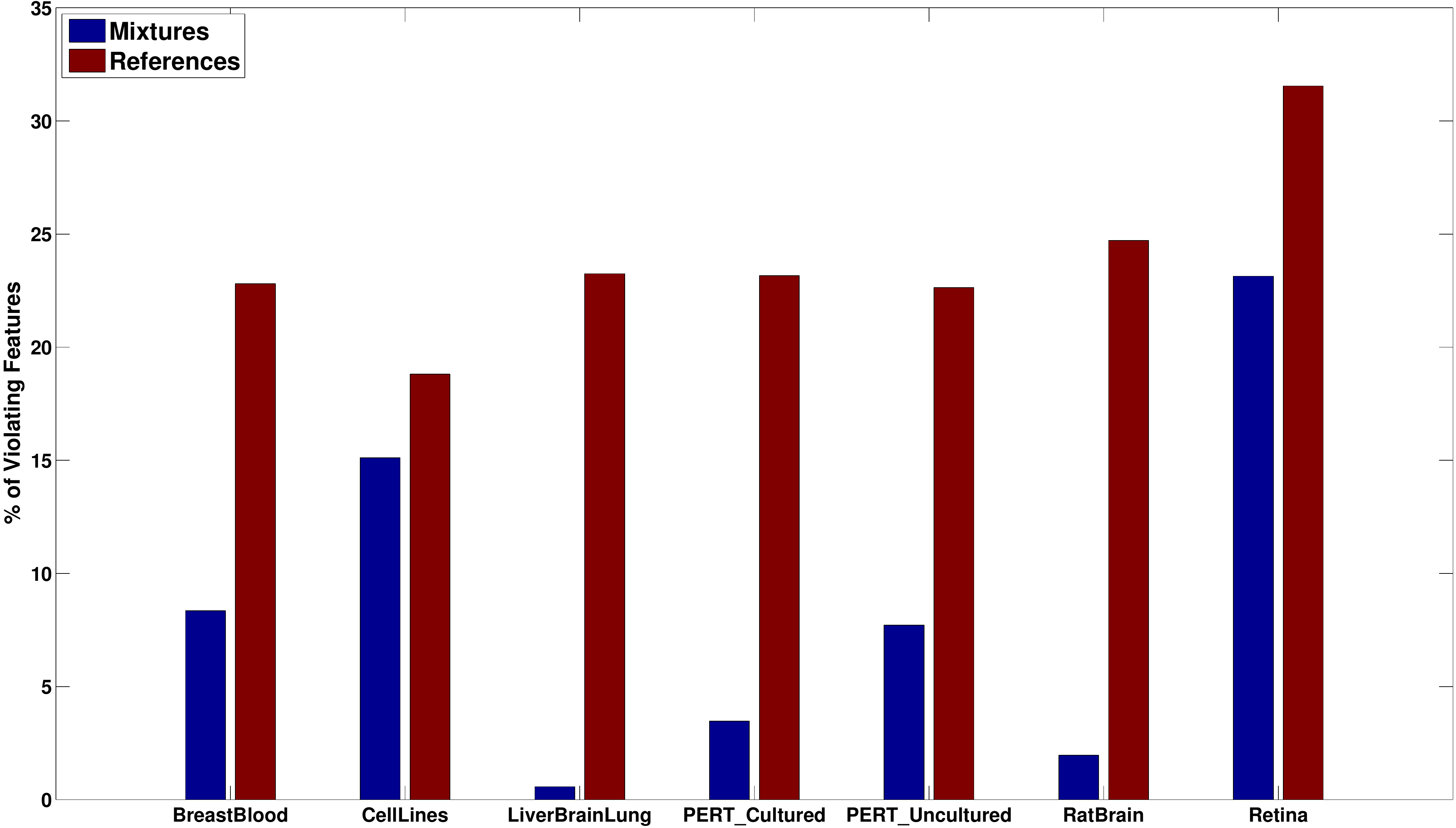}
\caption{Percent of features in each dataset that violate the STO constraint} 
\label{fig:STO_violation}
\end{figure}

Figure~\ref{fig:STO_violation} presents violating features in mixtures and reference 
profiles, averaged over all mixture samples in each dataset. We normalize and report the 
percent of features to account for differences in the total number of features in each dataset. We 
first observe that for the majority of datasets, except \textbf{Retina} and 
\textbf{BreastBlood}, the percent of violating features is much smaller than violating features in 
reference profiles. These two datasets also have the highest number of violating features in 
their reference profiles, summing to a total of approximately $60\%$ of all features. This 
observation is likely due to the normalization used in pre-processing microarray 
profiles. Specifically, one must not only normalize $\Matrix{M}$ and $\Matrix{G}$ 
independently, but also with respect to each other. We suggest using control genes 
that are expressed in all cell-types with low variation to normalize expression profiles. 
A recent study aimed to identify subsets of housekeeping genes in human tissues that respect
these conditions \cite{Eisenberg2013}. Another choice is using ribosomal proteins, the
basic building blocks of the cellular translation machinery, which are expressed in
a wide range of species. The Remove Unwanted Variation (RUV) \cite{Gagnon-Bartsch2012}
method is developed to remove batch effects from microarray and RNASeq expression profiles,
but also to normalize them using control genes. A simple extension of this method can
be adopted to solve the normalization difference between mixtures and references.

\begin{figure}[!t]
\centering
\includegraphics[width=0.8\columnwidth]{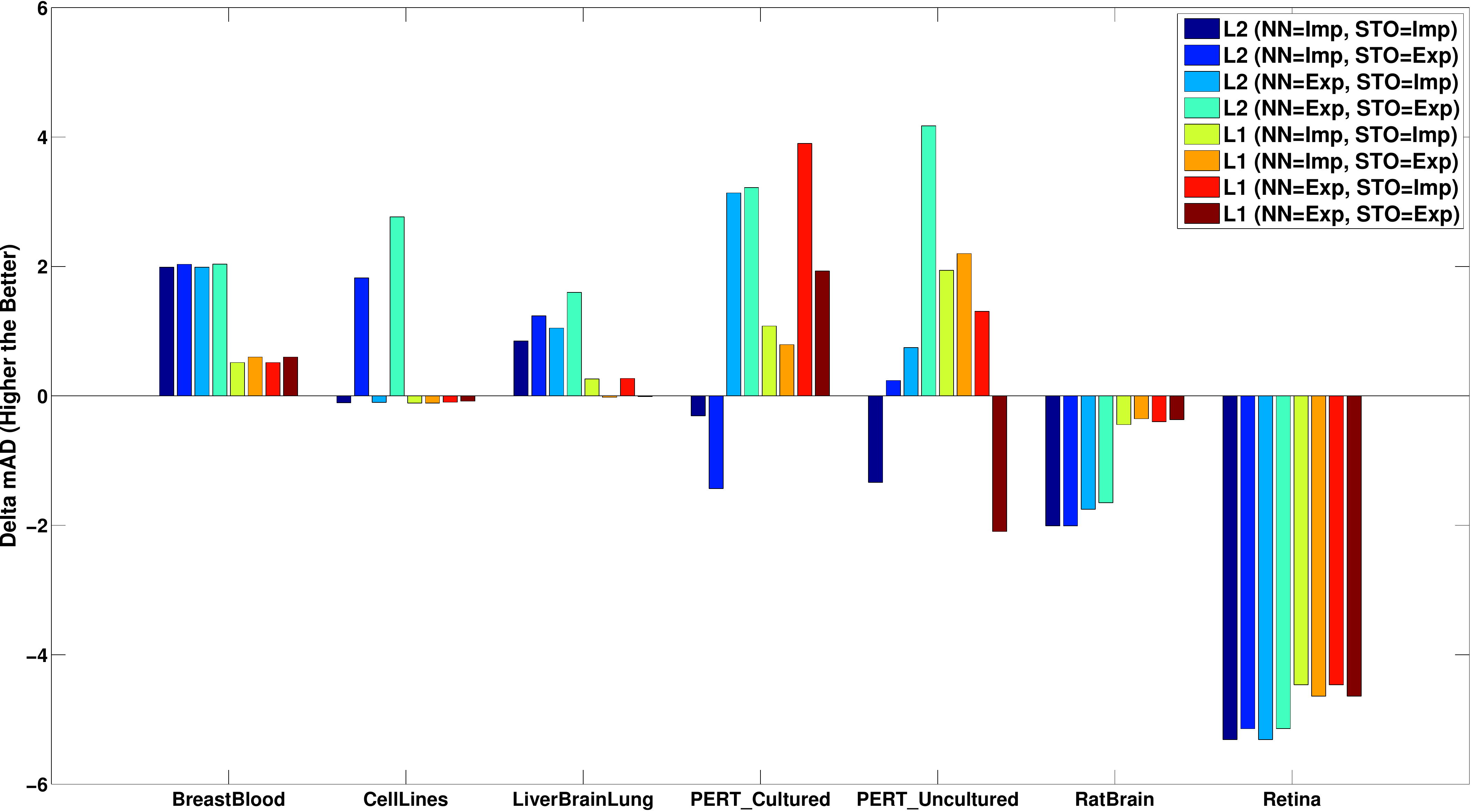}
\caption{Performance of deconvolution methods after removing violating features} 
\label{fig:delta_mAD_STO}
\end{figure}

Next, we evaluate how filtering these features affects
deconvolution performance of each dataset. For each case, we run deconvolution using all 
configurations and report the change (delta mAD) independently. 
Figure~\ref{fig:delta_mAD_STO} presents changes in the $mAD$ estimation error after removing 
violating features in both $\Vector{m}$ and $\Matrix{G}$ before performing deconvolution. 
Similar to previous experiments, the \textbf{Retina} dataset exhibits widely different behavior 
than the rest of the datasets. Removing this dataset from further consideration, we find that the 
overall performance over all datasets improves, with the exception of the \textbf{RatBrain} dataset. In the case 
of the \textbf{RatBrain} dataset, we hypothesize that the initially superior performance can be 
attributed to highly expressed features. These outliers, that happens to agree with the true solution, result
in \textit{over-fitting}. Finally, we note a correlation between observed enhancements and the 
level of violation of features in $\Vector{m}$. Consistent with this observation, we obtain
similar results when we only filter violating features from mixtures, but not reference 
profiles.

\subsection{Range Filtering-- Finding an Optimal Threshold}
\label{sec:range_filtering_effect}

Different upper/lower bounds have been proposed in the literature to prefilter expression
values prior to deconvolution. For example, Gong  \textit{et al.} \cite{Gong2011} suggest
an effective range of $[0.5, 5000]$, whereas Ahn \textit{et al.} \cite{Ahn2013} observe
an optimal range of $[2^4-2^{14}]$. To facilitate the choice of expression bounds, we 
seek a systematic way to identify an optimal range for different datasets. Kawaji \textit{et 
al.} \cite{Kawaji2014} recently report on an experiment to assess whether gene expression is 
quantified linearly in mixtures. To this end, they mix two cell-types (THP-1 and HeLa 
cell-lines) and see if experimentally measured expressions match with the 
computationally simulated datasets. They observe that expression values for microarray 
measurements are skewed for the lowly expressed genes (approximately $< 10$). This allows us 
to choose the lower bound based on experimental evidence. In our study, we search for
the optimal bounds over a $log_2$-linear space; thus, we set a threshold of $2^3$ on
the minimum expression values, which is closest to the bound proposed by Kawaji
\textit{et al.} \cite{Kawaji2014}.

Choosing an upper bound on the expression values is a harder problem, since it 
relates to enhancing the performance of deconvolution methods by removing outliers. 
Additionally, there is a known relationship between the mean expression value and its 
variance \cite{Tu2002}, which makes these outliers noisier than the rest of the features. This 
becomes even more important when dealing with purified cell-types that come from 
different labs, since highly expressed time/micro-environment dependent genes would be 
significantly different than the ones in the mixture \cite{Qiao2012}. A simple argument is to 
filter genes that the range of expression values in \textit{Affymetrix} microarray technology
is bounded by $2^{16}$ (due to initial normalization and image processing steps). 
Measurements close to this bound are not reliable as they might be saturated and inaccurate. 
However, practical bounds used in previous studies are far from these extreme values. In order to
examine the overall distribution of expression 
values, we analyze different datasets independently. For each dataset, we separately analyze
mixture samples and reference profiles, encoded by matrices $\Matrix{M}$ and $\Matrix{G}$, 
respectively. For each of these matrices, we vectorize the expression values and perform 
kernel smoothing using the Gaussian kernel to estimate the probability density function. 

\begin{figure}[!t]
\centering
\subfigure[Mixtures]{\includegraphics[width=.45\textwidth]{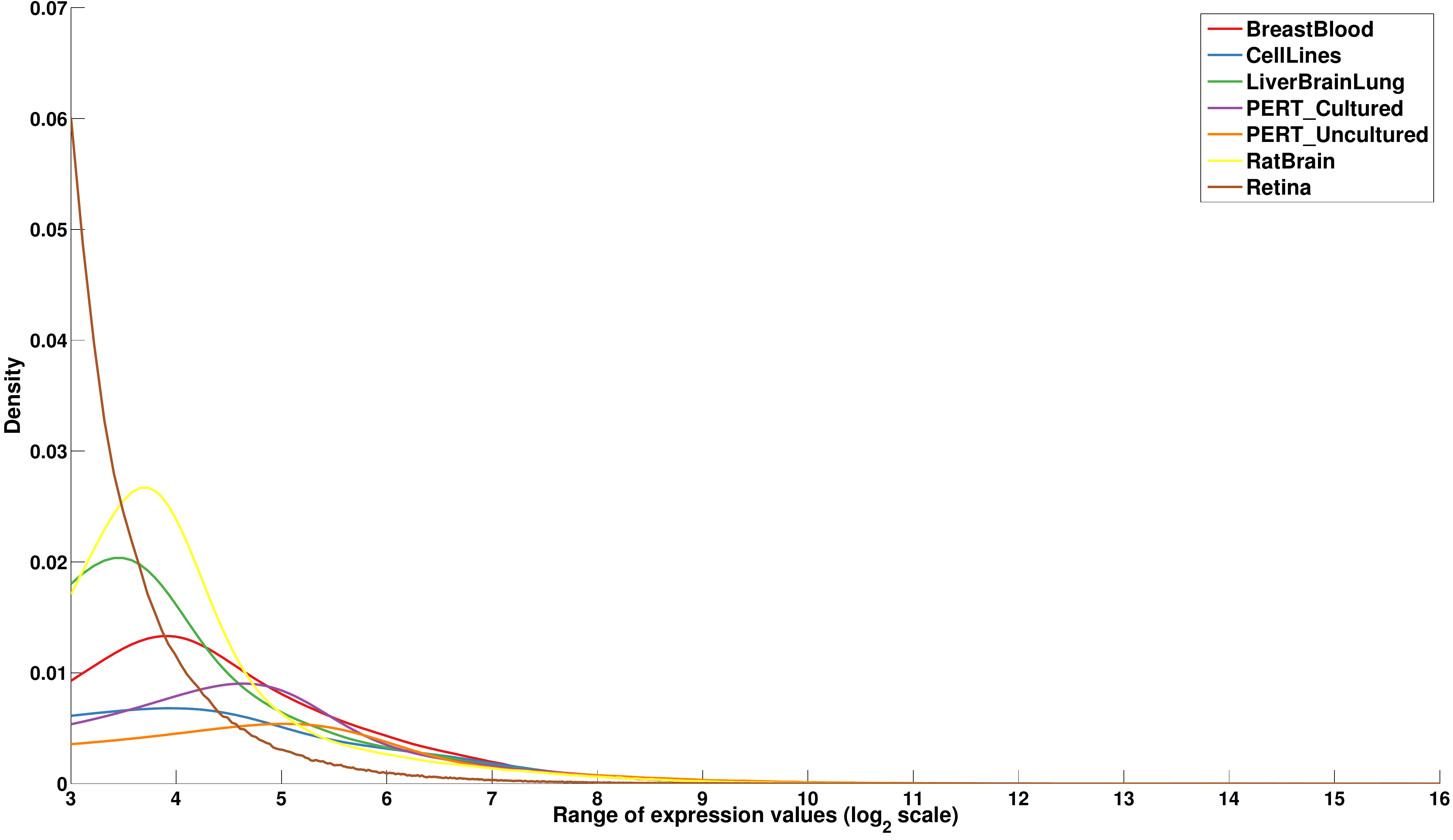}%
\label{fig:mix_dist}}
\hfil
\subfigure[Reference Profiles]{\includegraphics[width=.45\textwidth]{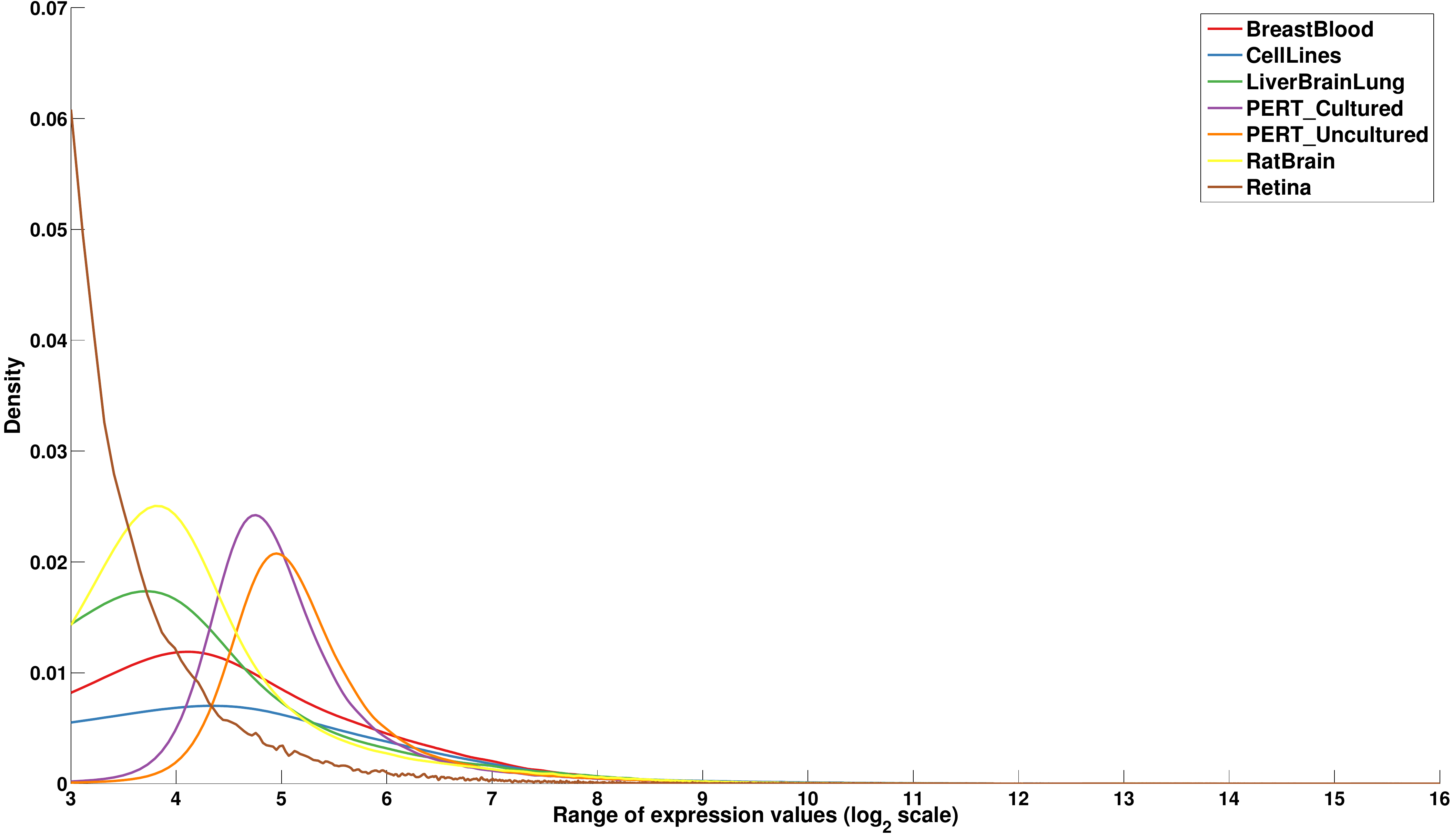}%
\label{fig:ref_dist}}
	\caption{Distribution of expression values}
\label{fig:expr_dist}
\end{figure}

Figure~\ref{fig:mix_dist} and Figure~\ref{fig:ref_dist} show the distribution of 
log-transformed expression values for mixtures and reference profiles, respectively. These 
expression values are greater than our lower bound of $2^3$. In agreement with our previous 
results, we observe an unusually skewed distribution for the \textbf{Retina} dataset, which in 
turn contributes to its lower performance compared to other ideal mixtures. Additionally, 
we observe that approximately $80\%$ of the features in this dataset are smaller than $2^3$, which 
are filtered and not shown in the distribution plot. For the rest of the datasets, in both 
mixtures and references, we observe a bell-shaped distribution with most of the features 
captured up to an upper bound of $2^{8}-2^{10}$. Another exception to this pattern is the 
\textbf{CellLines} dataset, which has a heavier tail than other datasets, especially in its 
reference profile.

\begin{figure}[!t]
\centering
\includegraphics[width=0.8\columnwidth]{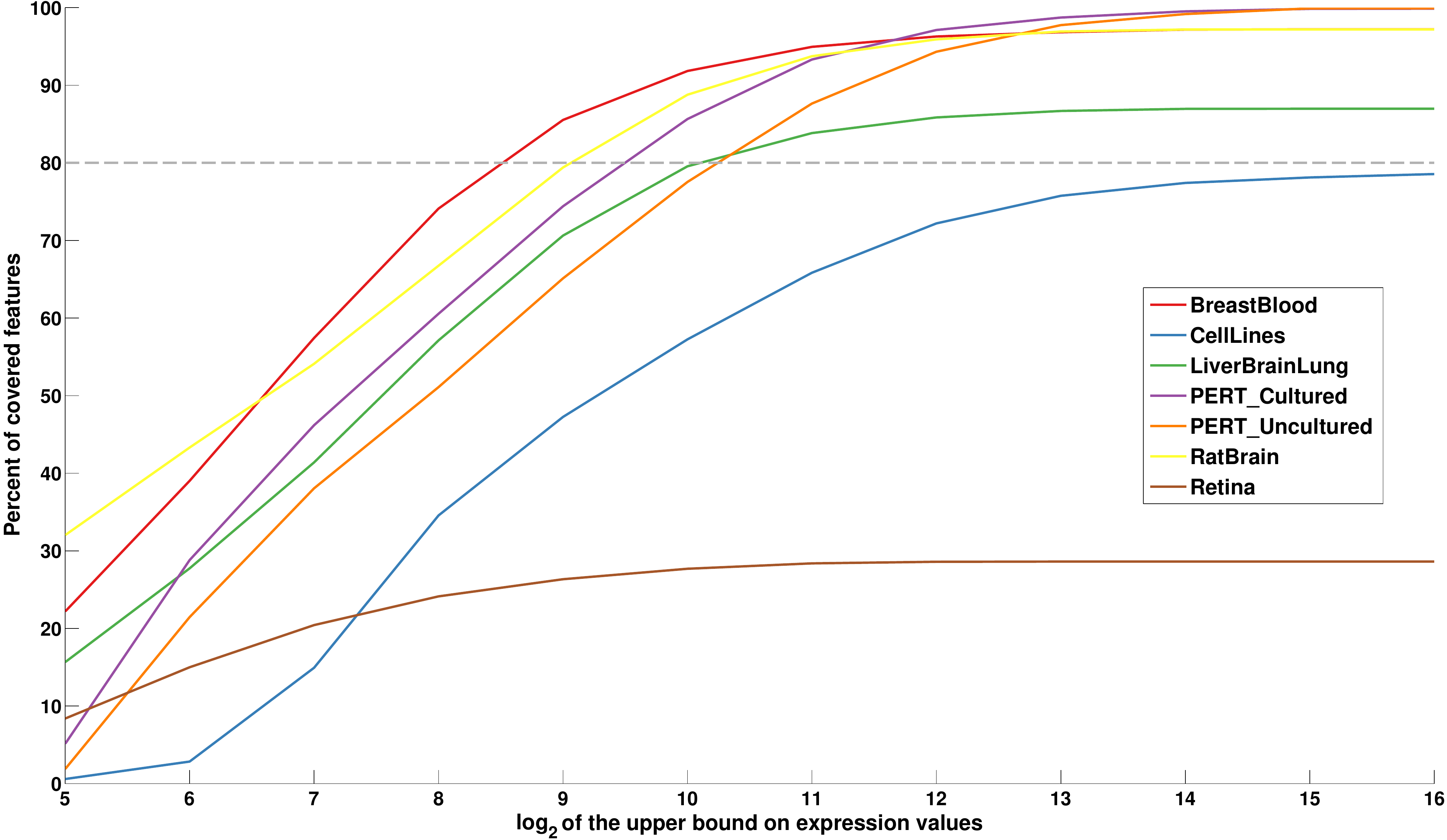}
\caption{Percent of covered features during range filtering} 
\label{fig:exp3_remainingFeatures}
\end{figure}

Next, we systematically evaluate the effect of range filtering by analyzing upper
bounds increasing in factors of 10 in the range $\{ 2^5, \cdots, 2^{16}\}$. In each case,
we remove all features that at least one of 
the reference profiles or mixture samples has a value exceeding this upper bound. 
Figure~\ref{fig:exp3_remainingFeatures} illustrates the percent of features that are retained,
as we increase the upper bound. As mentioned earlier, approximately $80\%$ of the features in 
the \textbf{Retina} dataset are lower than $2^3$, which is evident from the maximum percent of 
features left to be bounded by $20\%$ in this figure. Additionally, consistent with our 
previous observation over expression densities, more that $80\%$ of the features are covered 
between $2^{8}-2^{10}$, except for the \textbf{CellLine} dataset.

\begin{figure}[!t]
\centering
\subfigure[Cultured]{\includegraphics[width=.45\textwidth]{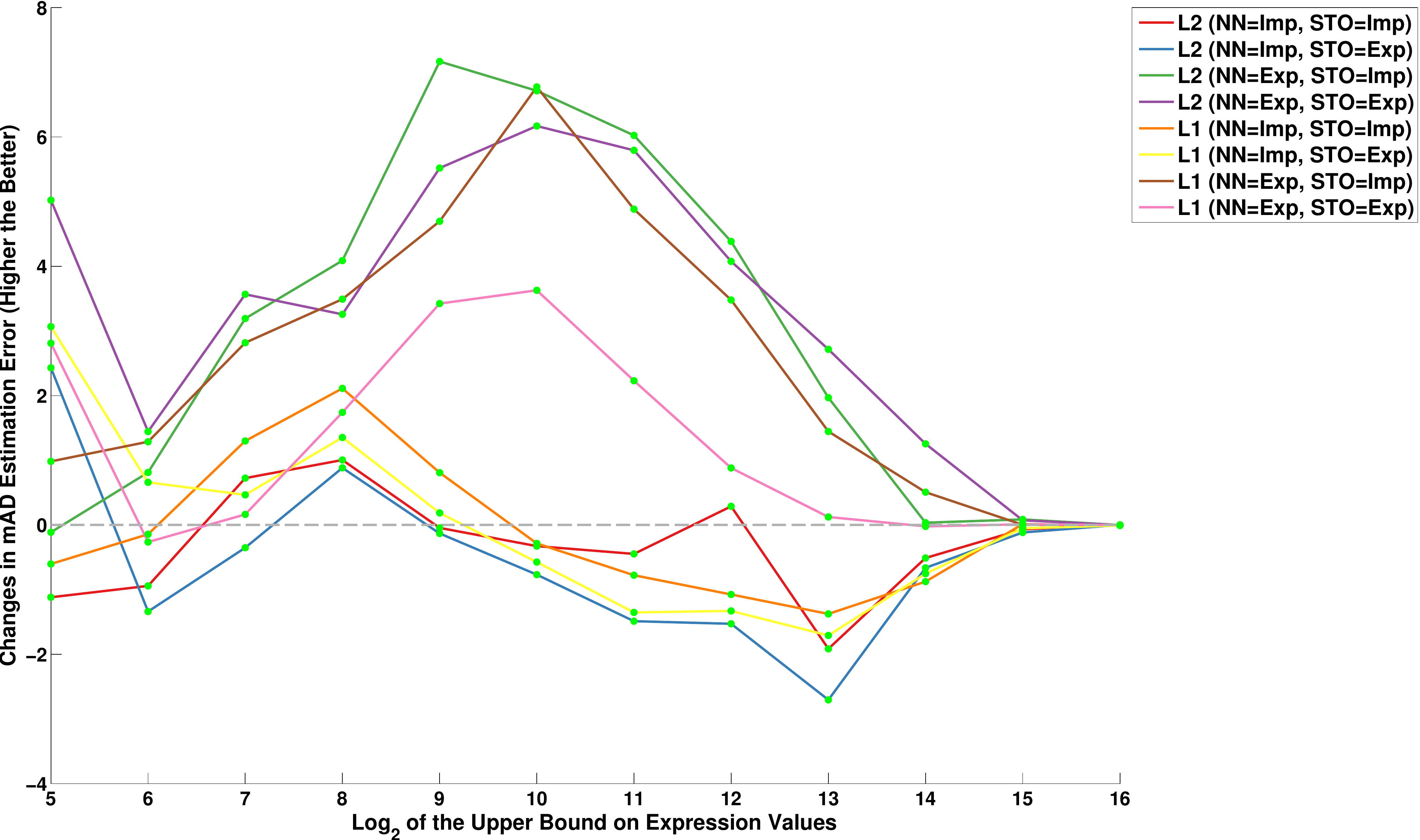}%
\label{fig:exp3_pertCult}}
\hfil
\subfigure[Uncultured]{\includegraphics[width=.45\textwidth]{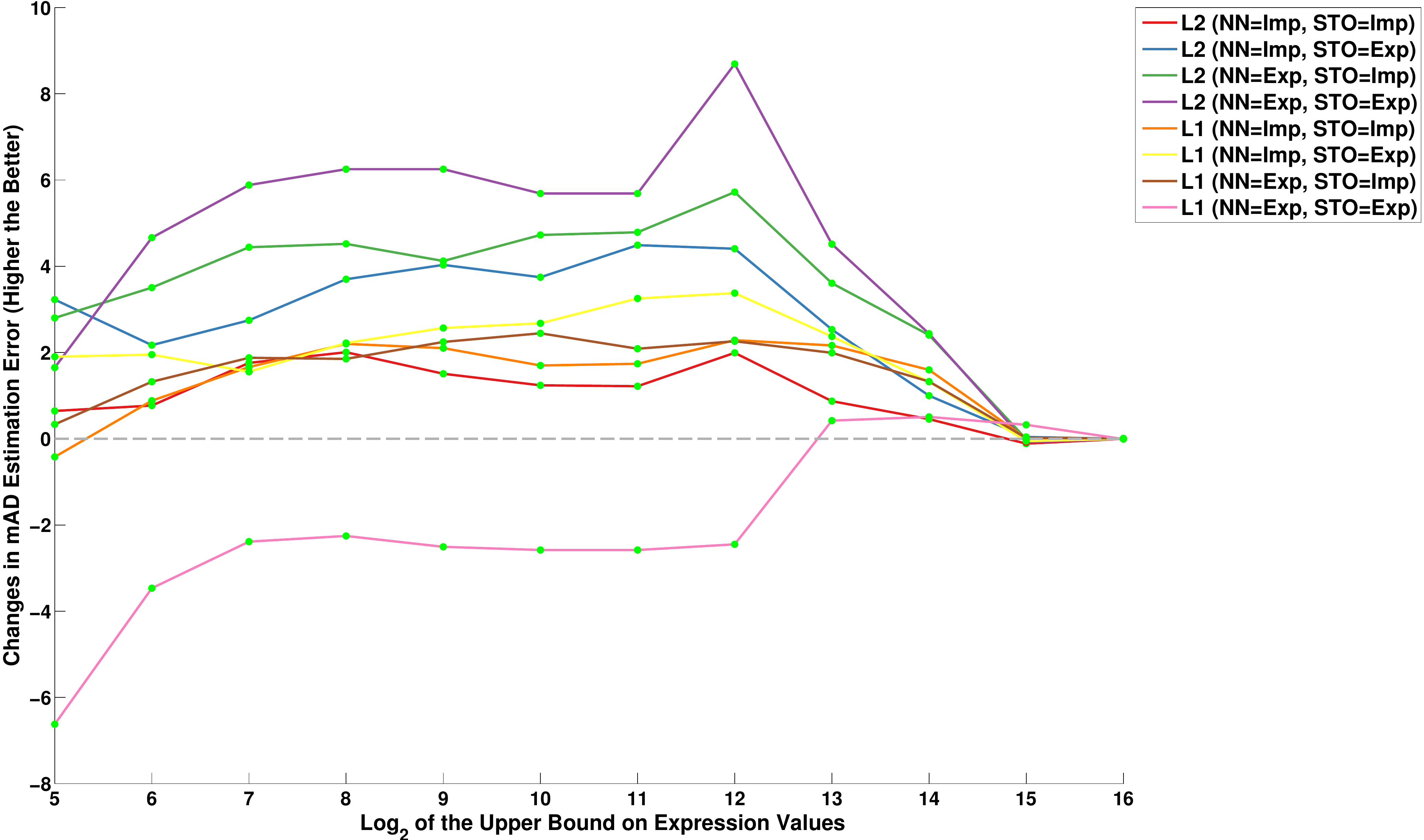}%
\label{fig:exp3_pertUnCult}}
	\caption{Performance of PERT datasets during range filtering}
\label{fig:expr_dist}
\end{figure}

Finally, we perform deconvolution using the remaining features given each upper bound. The 
results are mixed, but a common trend is that removing highly expressed genes decreases 
performance of ideal mixtures with known concentrations, while enhancing the performance of 
\textbf{PERT} datasets. Figure~\ref{fig:exp3_pertCult} and Figure~\ref{fig:exp3_pertUnCult} 
show the changes in mAD error, compared to unfiltered deconvolution, for the \textbf{PERT} 
dataset. In each case, we observe improvements up to 7 and 8 percent, respectively. The red 
and green points on the diagram show the significance of deconvolution. Interestingly, while 
both methods show similar improvements, all data points for cultured PERT seem to be 
insignificant, whereas uncultured PERT shows significance for the majority of data-points. 
This is due to the weakness of our random model, which is dependent on the number of samples and 
is not comparable across datasets. Uncultured PERT has twice as many samples as cultured PERT, 
which makes it less likely to have any random samples achieving as good an mAD as the observed 
estimation error. This dependency on the number of samples can be addressed by defining 
sample-based \emph{p}-values. Another observation is that for the uncultured dataset, all measures 
have been improved, except $\Loss_1$ with explicit \textit{NN} and \textit{STO} constraints. 
On the other hand, for the cultured dataset, both $\Loss_1$ and $\Loss_2$ with the explicit 
\textit{NN} constraint perform well, whereas implicitly enforcing \textit{NN} deteriorates
their performance. Cultured and uncultured datasets have their peak at $2^{10}$ and $2^{12}$, 
respectively.

\begin{figure}[!t]
\centering
\includegraphics[width=0.8\columnwidth]{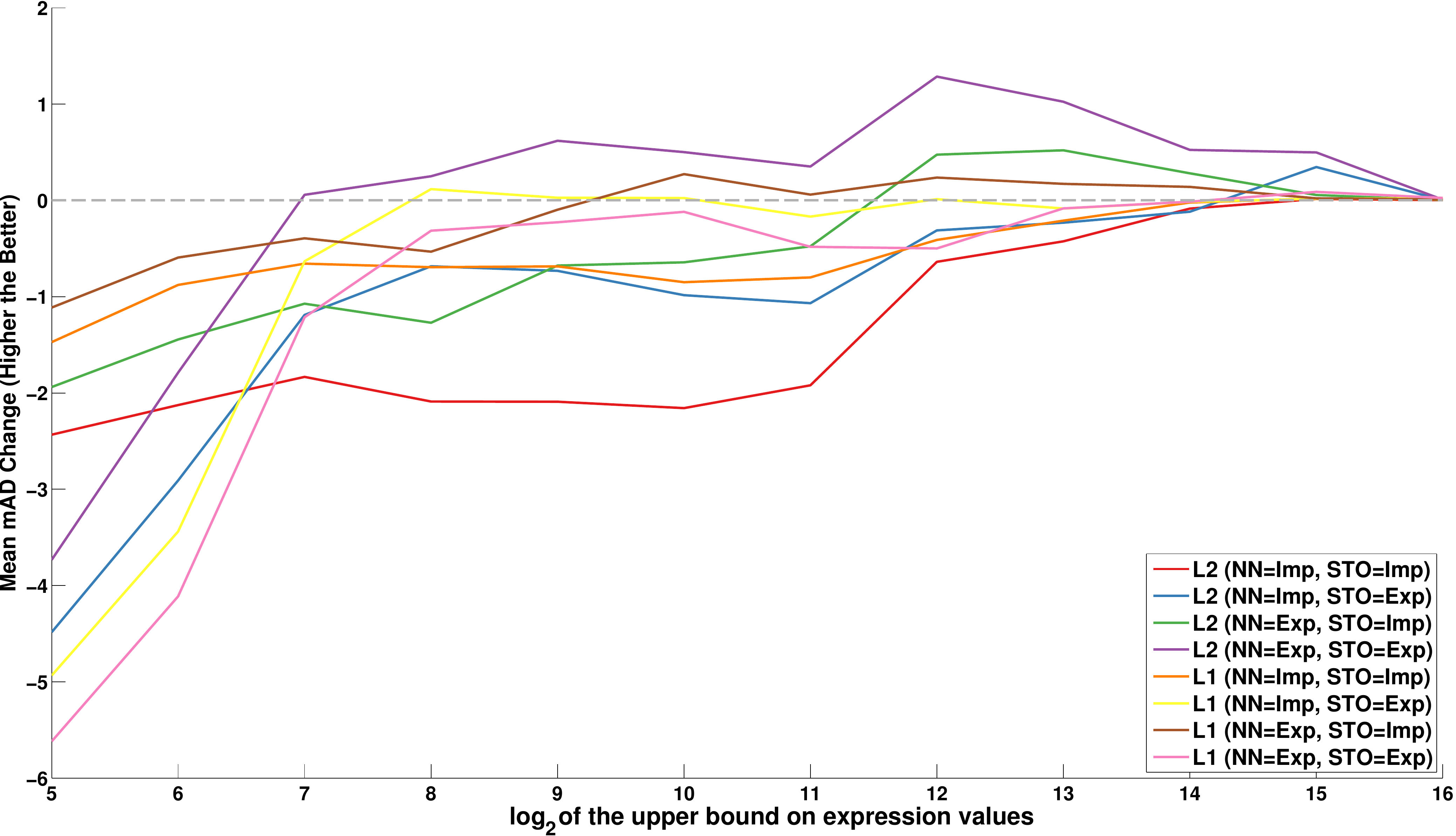}
\caption{Average performance of range filtering over all datasets} 
\label{fig:exp3_avgPerf}
\end{figure}

For the rest of the datasets, range filtering decreased performance in a majority of 
cases, except the \textbf{Retina} dataset, which had an improved performance at $2^{6}$ with the 
best result achieved with $\Loss_1$ with both explicit \textit{NN} and \textit{STO} 
enforcement. This changed the best observed performance of this datasest, measured as mAD, to be 
close to 7. These mixed results make it harder to choose a threshold for the upper bound, so 
we average results over all datasets to find a balance between improvements in PERT and 
overall deterioration in other datasets. Figure~\ref{fig:exp3_avgPerf} presents the averaged 
mAD difference across all datasets. This suggests a ``general" upper bound filter of $2^{12}$ 
to be optimal across all datasets. 

\begin{figure}[!t]
\centering
\includegraphics[width=0.8\columnwidth]{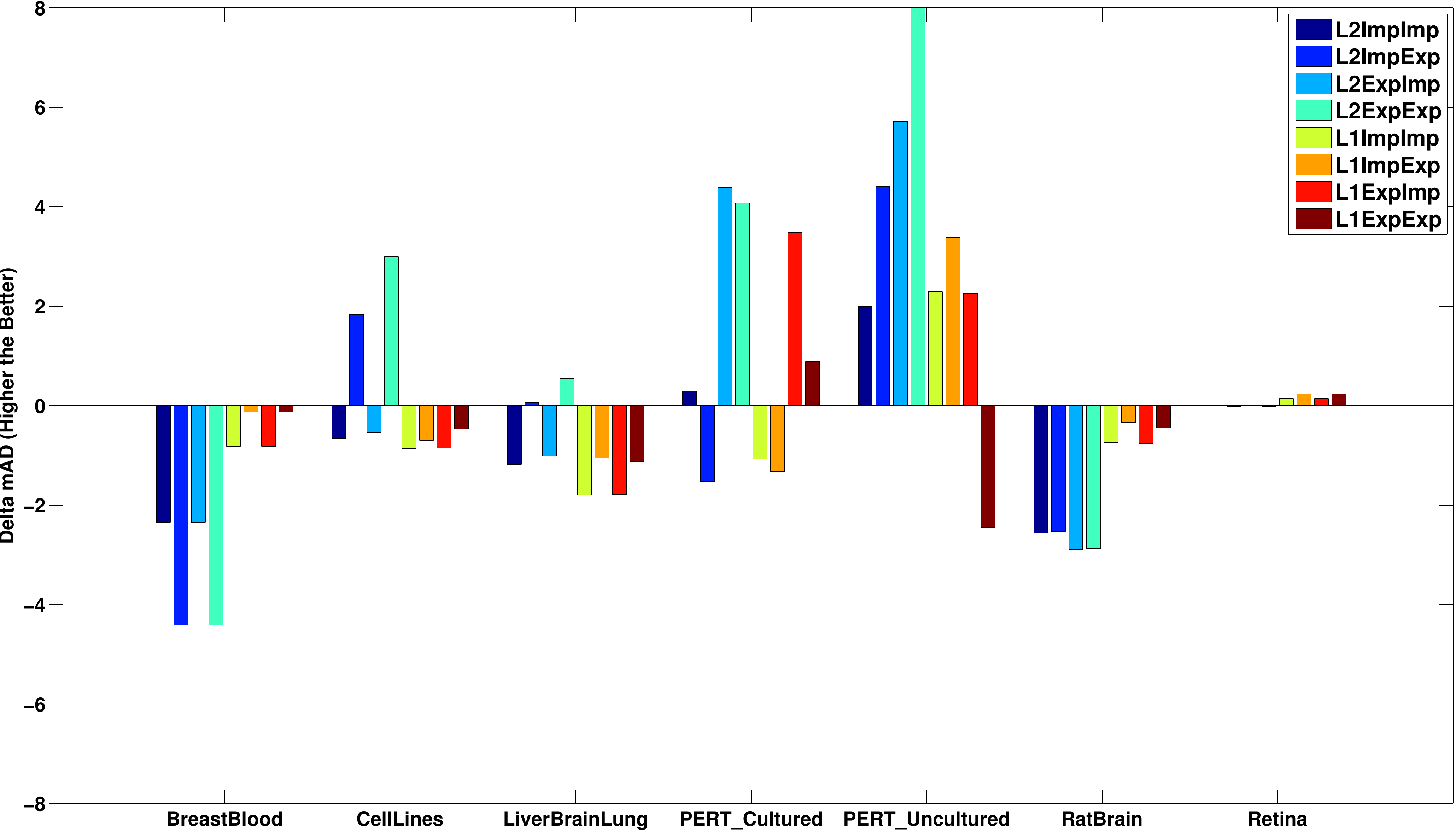}
\caption{Dataset-specific changes in the performance of deconvolution methods after filtering expression ranges to fit within $[2^3-2^{12}]$} 
\label{fig:fixed_range_performance}
\end{figure}

We use this threshold to filter all datasets and perform deconvolution on them.
Figure~\ref{fig:fixed_range_performance} presents the dataset-specific performance
of range filtering with fixed bounds, measured by changes in the \textit{mAD} value
compared to the original deconvolution. As observed from individual performance plots,
range filtering is most effective in cases where the reference profiles differ
significantly from the true cell-types in the mixture, such as the case with
the \textbf{PERT} datasets. In ideal mixtures, since cell-types are measured and
mixed at the same time/laboratory, this distinction is negligible. In these cases,
highly expressed genes in mixtures and references coincide with each other and
provide additional clues for the regression. Consequently, removing these highly
expressed genes often degrades the performance of deconvolution methods.
This generalization of the upper bound threshold, however, should be adopted with care,
since each dataset exhibits different behavior in response to range filtering.
Ideally, one must filter each dataset individually based on the distribution of
expression values. Furthermore, in practical applications, gold standards are not
available to aid in the choice of cutoff threshold.

\begin{figure}[!t]
\centering
\includegraphics[width=0.8\columnwidth]{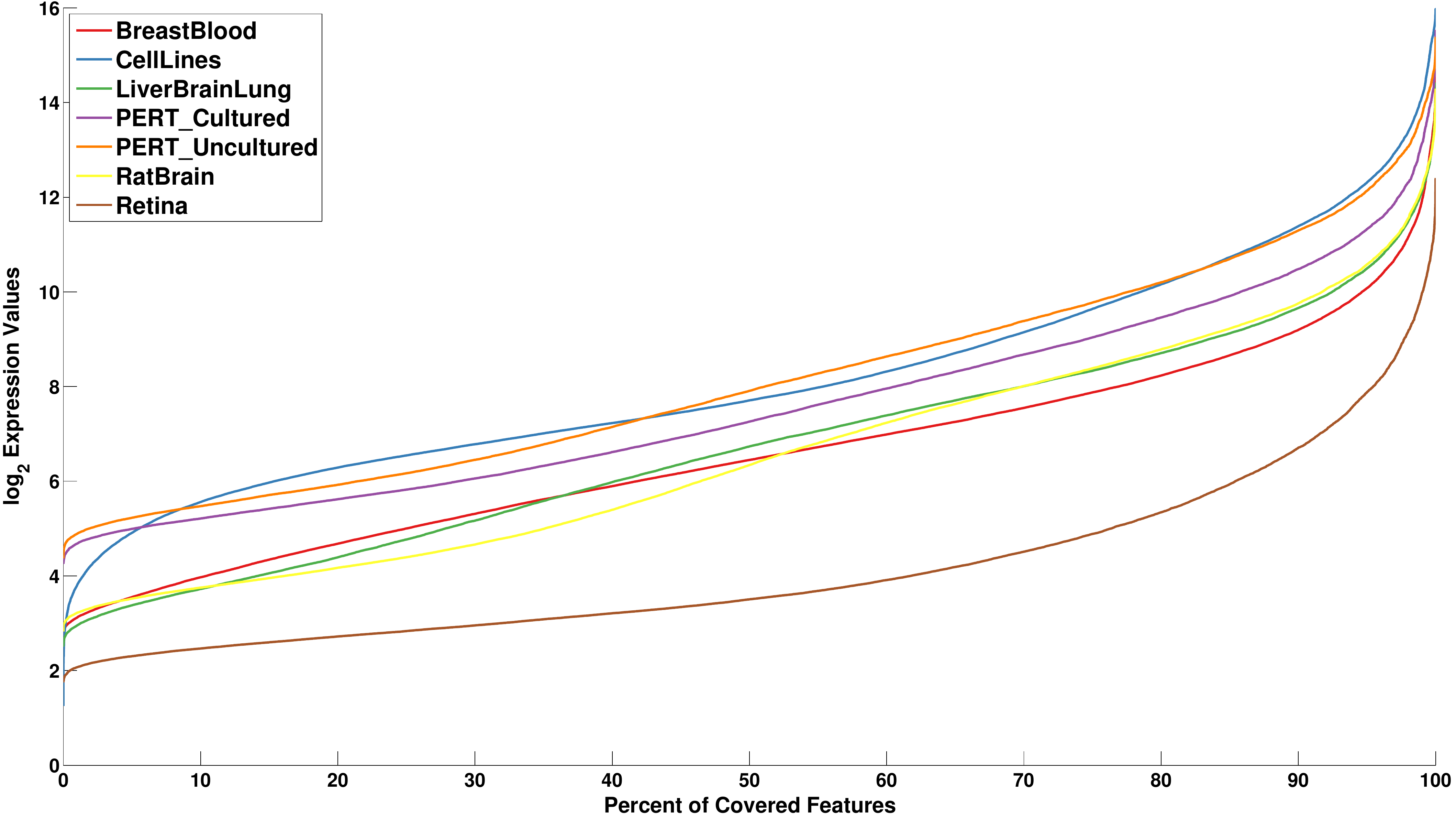}
\caption{Sorted $log_2$-transformed gene expressions in different datasets} 
\label{fig:sorted_exp}
\end{figure}

\begin{figure}[!t]
\centering
\includegraphics[width=0.8\columnwidth]{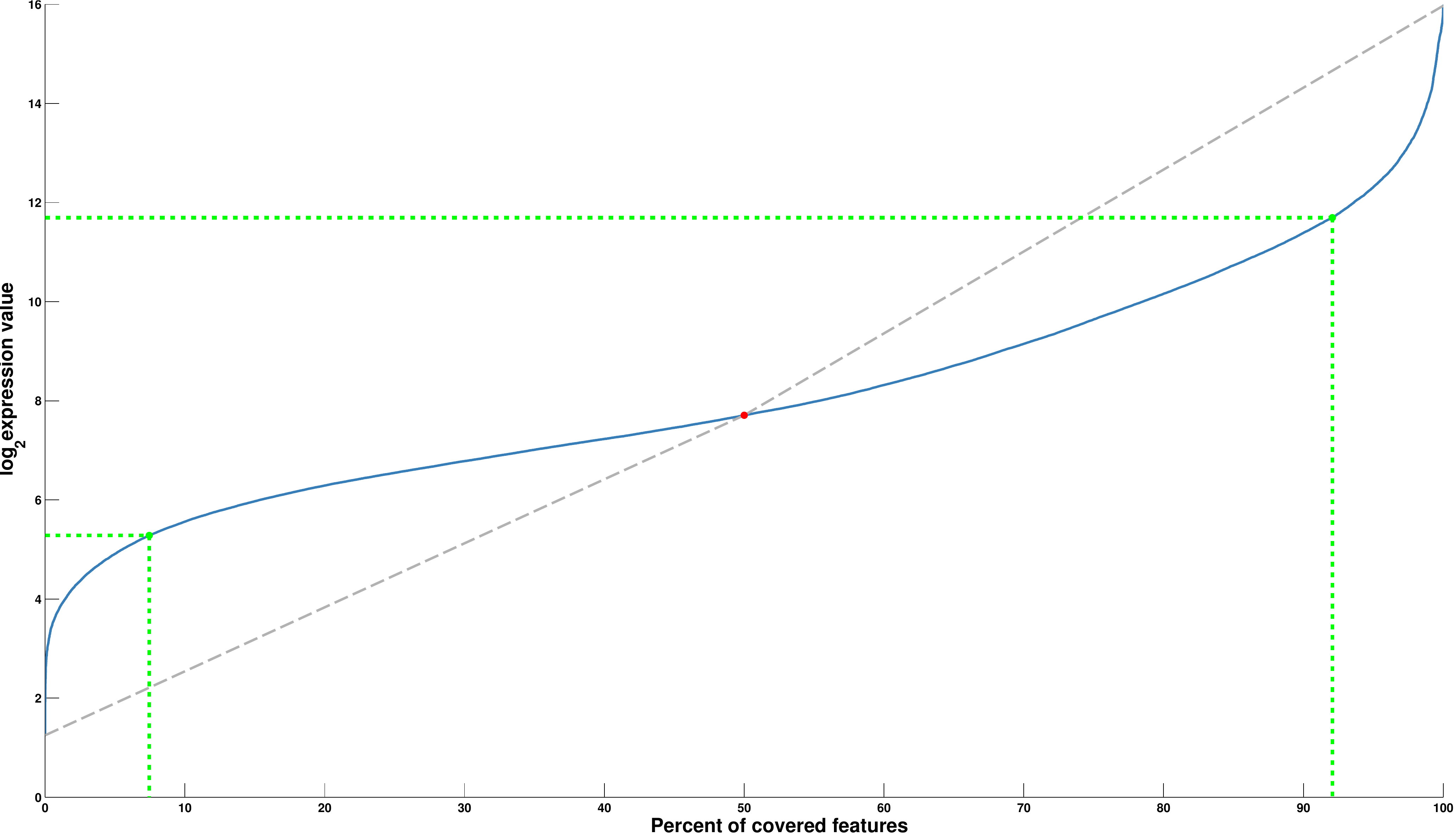}
\caption{Example of adaptive filtering over \textbf{CellLines} dataset} 
\label{fig:CC_example}
\end{figure}

We now introduce a new method that adaptively identifies an effective range for each dataset.
Figure~\ref{fig:sorted_exp} illustrates the $log_2$ normalized value of maximal expression
for each gene in matrices $\Matrix{M}$ and $\Matrix{G}$, sorted in ascending order.
In all cases, intermediate values exhibit a gradual increase, whereas the top and bottom
elements in the sorted list show a steep change in their expression. We aim to identify
the critical points corresponding to these sudden changes in the expression values for
each dataset. To this end, we select the middle point as a point of reference and
analyze the upper and lower half, independently. For each half, we find the point on
the curve that has the longest distance from the line connecting the first (last)
element to the middle element. Application of this process over the \textbf{CellTypes}
dataset is visualized in Figure~\ref{fig:CC_example}. Green points in this figure
correspond to the critical points, which are used to define the lower and upper bound
for the expression values of this dataset.

\begin{figure}[!t]
\centering
\includegraphics[width=0.8\columnwidth]{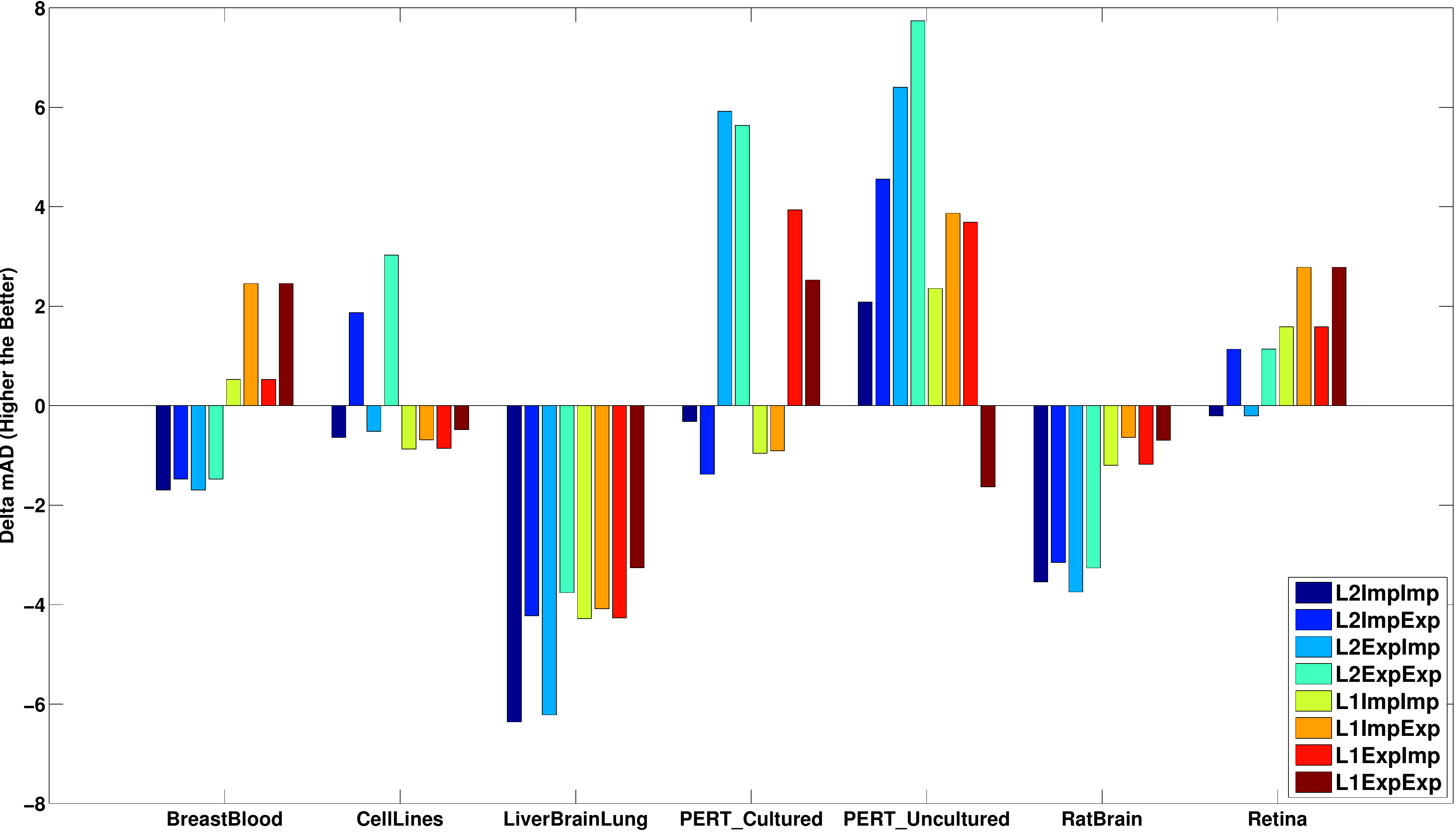}
\caption{Dataset-specific changes in the performance of deconvolution methods after adaptive range filtering} 
\label{fig:adaptive_range_performance}
\end{figure}

\begin{figure}[!t]
\centering
\subfigure[LiverBrainLung]{\includegraphics[width=.45\textwidth]{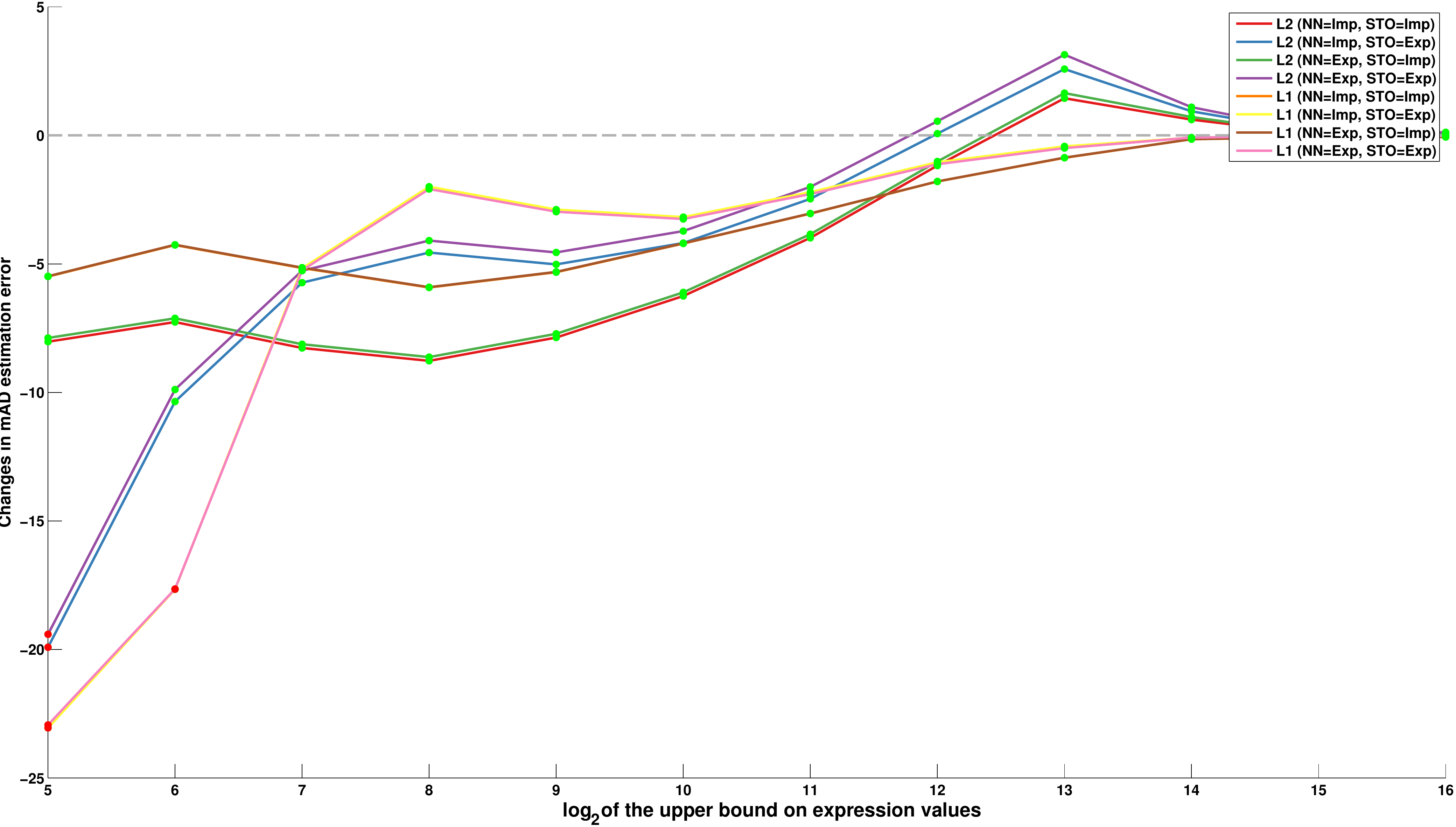}%
\label{fig:Liver_range_filtering}}
\hfil
\subfigure[RatBrain]{\includegraphics[width=.45\textwidth]{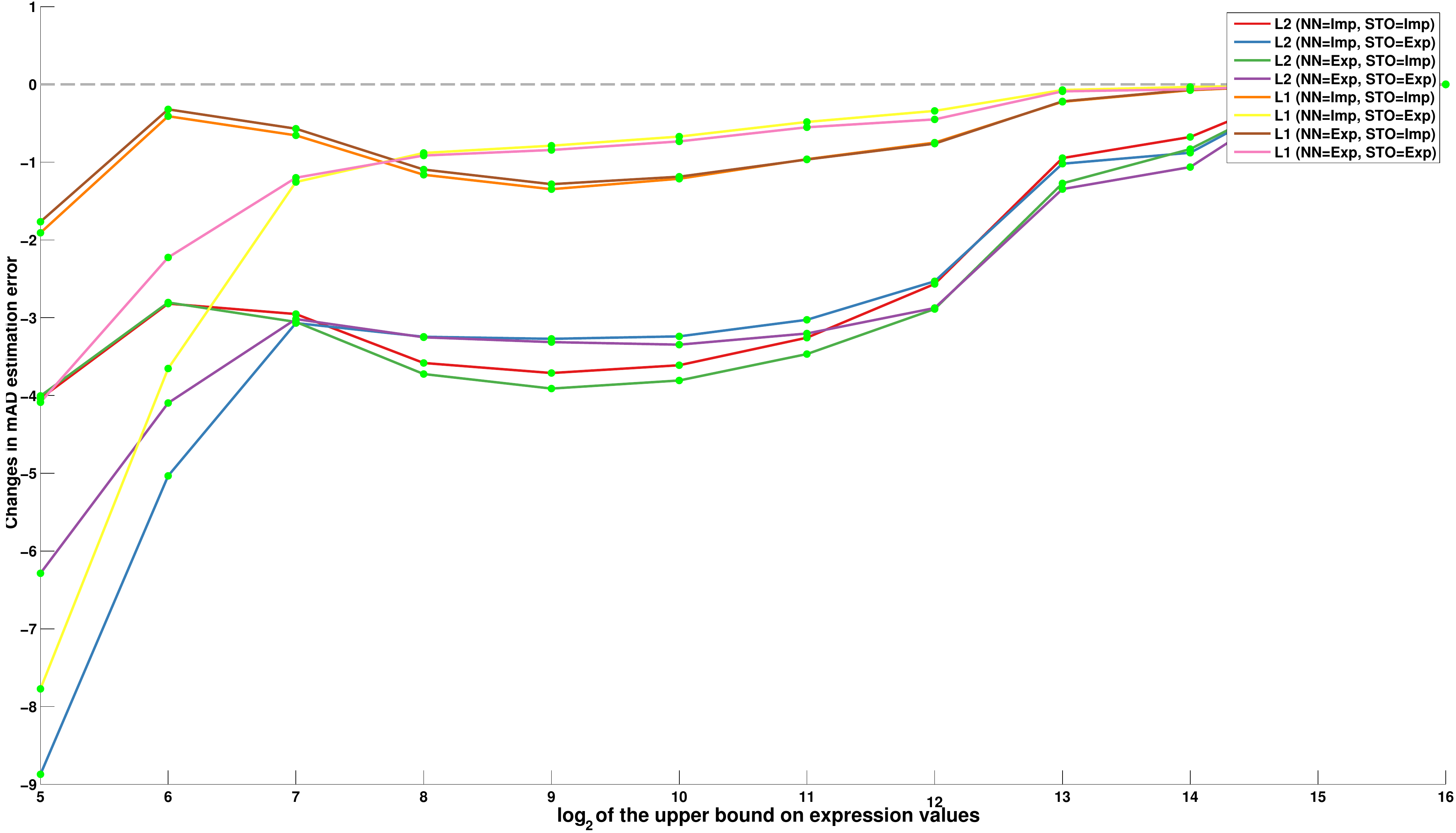}%
\label{fig:Rat_range_filtering}}
	\caption{Individual performance plots for range filtering in datasets which range filtering exhibits negative effect on the deconvolution}
\label{fig:bad_datasets_for_adaptive}
\end{figure}

We use this technique to identify adaptive ranges for each dataset prior to deconvolution.
Table~\ref{table:adaptive_ranges} summarizes the identified critical points for each
dataset. Figure~\ref{fig:adaptive_range_performance} presents the dataset-specific
performance of each method after adaptive range filtering. While in most cases the
results for fixed and adaptive range filtering are compatible, and in some cases
adaptive filtering gives better results, the most notable difference is the degraded
performance of \textbf{LiverBrainLung}, and, to some extent, \textbf{RatBrain} datasets.
To further investigate this observation, we examine individual experiments for these
datasets for fixed thresholds. Figure~\ref{fig:bad_datasets_for_adaptive}
illustrates individual plots for each dataset. The common trend here is that in
both cases range filtering, in general, degrades the performance of deconvolution
methods for all configurations. In other words, extreme values in these datasets
are actually helpful in guiding the regression, and any filtering negatively impacts
performance. This suggests that range filtering, in general, is not always helpful in
enhancing the deconvolution performance, and in fact in some cases;
for example the ideal datasets such as \textbf{LiverBrainLung}, \textbf{RatBrain}, and
\textbf{BreastBlood}; it can be counterproductive.

\begin{table}[!t]
\caption{Summary of adaptive ranges for each dataset}
\label{table:adaptive_ranges}
\centering
\begin{tabular}{l|cc}
	&LowerBound	&UpperBound\\\hline
BreastBlood	&4.2842	&9.4314\\
CellLines	&5.2814	&11.6942\\
LiverBrainLung	&3.3245	&9.9324\\
PERT\_Cultured	&4.9416	&10.9224\\
PERT\_Uncultured	&5.1674	&11.5042\\
RatBrain	&3.3726	&9.9698\\
Retina	&2.4063	&6.7499\\
\end{tabular}
\end{table}

\subsection{Selection of Marker Genes -- The Good, Bad, and Ugly}
\label{sec:marker_selection_effect}

Selecting marker genes that uniquely identify a certain tissue or cell-type, prior to deconvolution,
can help in improving the conditioning of matrix $\Matrix{G}$, thus
improving its discriminating power and stability of results, as well as decreasing the 
overall computation time. A key challenge in identifying ``marker" genes is the choice of 
method that is used to assess selectivity of genes. Various parametric and nonparametric 
methods have been proposed in literature to identify differentially expressed genes 
between two groups \cite{Jeanmougin2010, Clark2014} or between a group and other 
groups \cite{VanDeun2009}. Furthermore, different methods have been developed in parallel to 
identify \textit{tissue-specific} and \textit{tissue-selective} genes that are unique markers 
with high specificity to their host tissue/cell 
type \cite{Cavalli2011,Kadota2006,Birnbaum2011, Mohammadi2015}. While choosing/developing 
accurate methods for identifying reliable markers is an important challenge, an in-depth 
discussion of the matter is beyond the scope of this article. Instead, we adopt two 
methods used in the literature. Abbas \textit{et al.} \cite{Abbas2009} present a framework 
for choosing genes based on their overall differential expression. For each gene, they use a
t-test to compare the cell-type with the highest expression with the second and third highest 
expressing cell-type. Then, they sort all genes and construct a sequence of basis matrices 
with increasing sizes. Finally, they use condition number to identify an ``optimal" cut among 
top-ranked genes that minimizes the condition number. Newman \textit{et 
al.} \cite{Newman2015} propose a modification to the method of Abbas 
\textit{et al.}, in which genes are not sorted based on their overall differential expression, 
but according to their tissue-specific expression when compared to all other cell-types. After 
prefiltering differentially expressed genes, they sort genes based on their expression fold 
ratio and use a similar cutoff that optimizes the condition number. Note that the
former method increases the size of the basis matrix by one at each step, while the latter method 
increases it by $q$ (number of cell-types). The method of Newman \textit{et 
al.} has the benefit that it chooses a similar number of markers per cell-type, which is useful in cases where one of 
the references has a significantly higher number of markers.

\begin{figure}[!t]
\centering
\includegraphics[width=0.8\columnwidth]{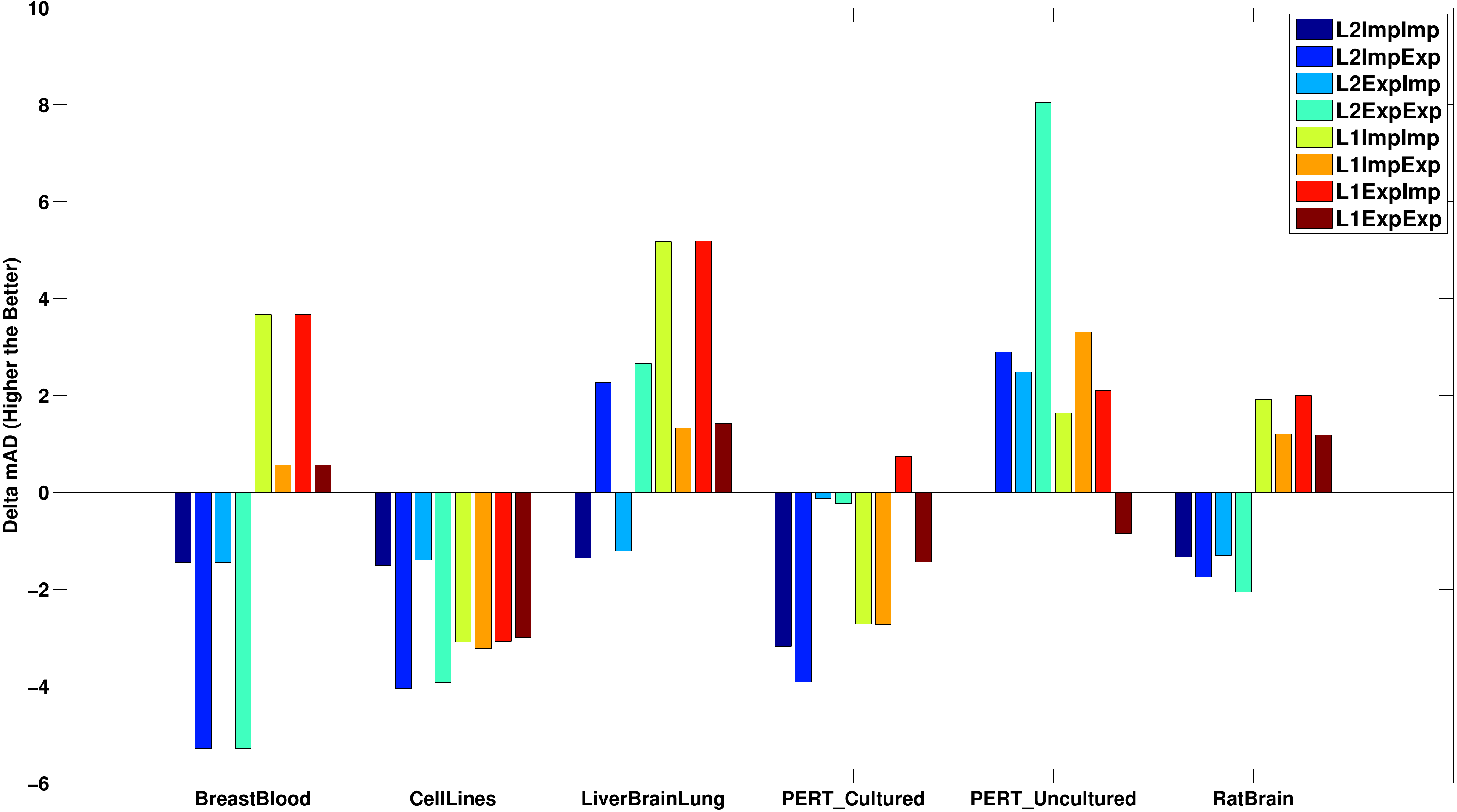}
\caption{Effect of marker selection on the performance of deconvolution methods} 
\label{fig:exp4_unlimitedMarkers}
\end{figure}

We implement both methods and assess their performance over the datasets. We observe slightly 
better performance with the second method and use it for the rest of our experiments. Due to 
unexpected behavior of the \textbf{Retina} dataset, as well as a low number of significant markers 
in all our trials, we eliminate this dataset from further study. In identifying differentially expressed genes,
a key parameter is the \emph{q}-value cutoff to report significant features.
The distribution of corrected \emph{p}-values exhibits high similarity among ideal 
mixtures, while differing significantly in \textbf{CellLines} mixtures and both \textbf{PERT} 
datasets. We find the range of $10^{-3}-10^{-5}$ to be an optimal balance between these two 
cases and perform experiments to test different cutoff values. 
Figure~\ref{fig:exp4_unlimitedMarkers} shows changes in the $mAD$ measure after applying 
marker detection, using a \emph{q}-value cutoff of $10^{-3}$, which resulted in the best 
overall performance in our study. We observe that the \textbf{PERT\_Uncultured} and 
\textbf{LiverBrainLung} datasets have the highest gain across the majority of configurations, 
while \textbf{BreastBlood} and \textbf{RatBrain} exhibit an improvement in experiments with 
$\Loss_1$ while their $\Loss_2$ performance is greatly decreased. Finally, for the 
\textbf{PERT\_Cultured} and \textbf{CellLines} datasets, we observe an overall decrease in 
performance in almost all configurations.

\begin{figure}[!t]
\centering
\includegraphics[width=0.8\columnwidth]{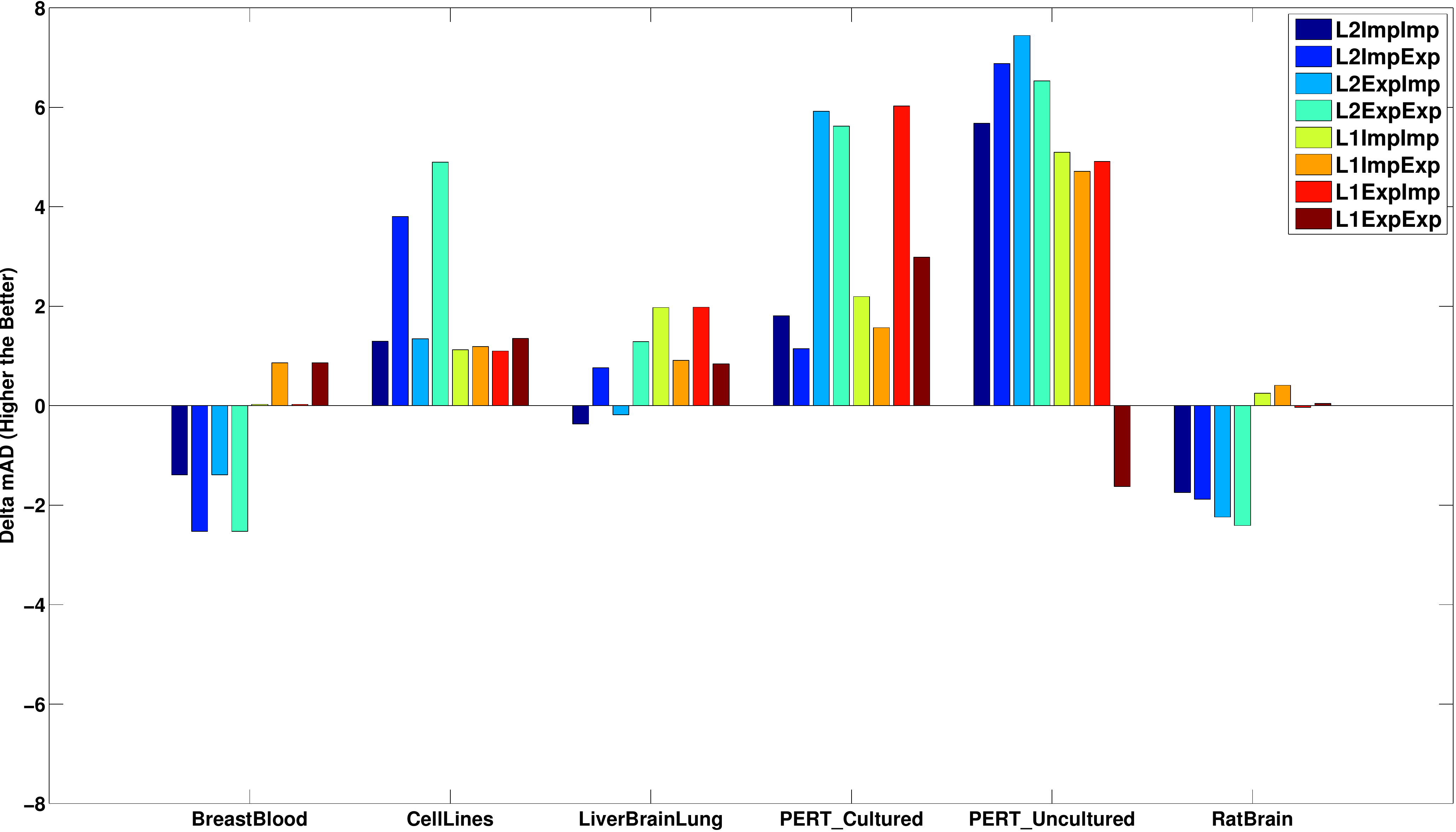}
\caption{Effect of marker selection, after range filtering, on the performance of deconvolution methods} 
\label{fig:exp4_limitedMarkers}
\end{figure}

Next, we note that the internal sorting based on fold-ratio intrinsically prioritizes highly 
expressed genes and is susceptible to noisy outliers. To test this hypothesis, we 
perform a range selection using a global upper bound of $10^{12}$ prior to the marker selection 
method and examine if this combination can enhance our previous results. We find the 
\emph{q}-value threshold of $10^{-5}$ to be the better choice in this case. 
Figure~\ref{fig:exp4_limitedMarkers} shows changes in performance of different methods 
when we prefilter expression ranges prior to marker selection. The most notable change is that 
both the \textbf{PERT\_Cultured} and the \textbf{CellLines}, which were among the datasets with the
lowest performance in the previous experiment, are now among the best-performing datasets, in 
terms of overall mAD enhancement. We still observe a higher negative impact on $\Loss_2$ in 
this case, but the overall amplitude of the effect has been dampened in both 
\textbf{BreastBlood} and \textbf{RatBrain} datasets.

\begin{figure}
\centering
\includegraphics[width=0.9\columnwidth]{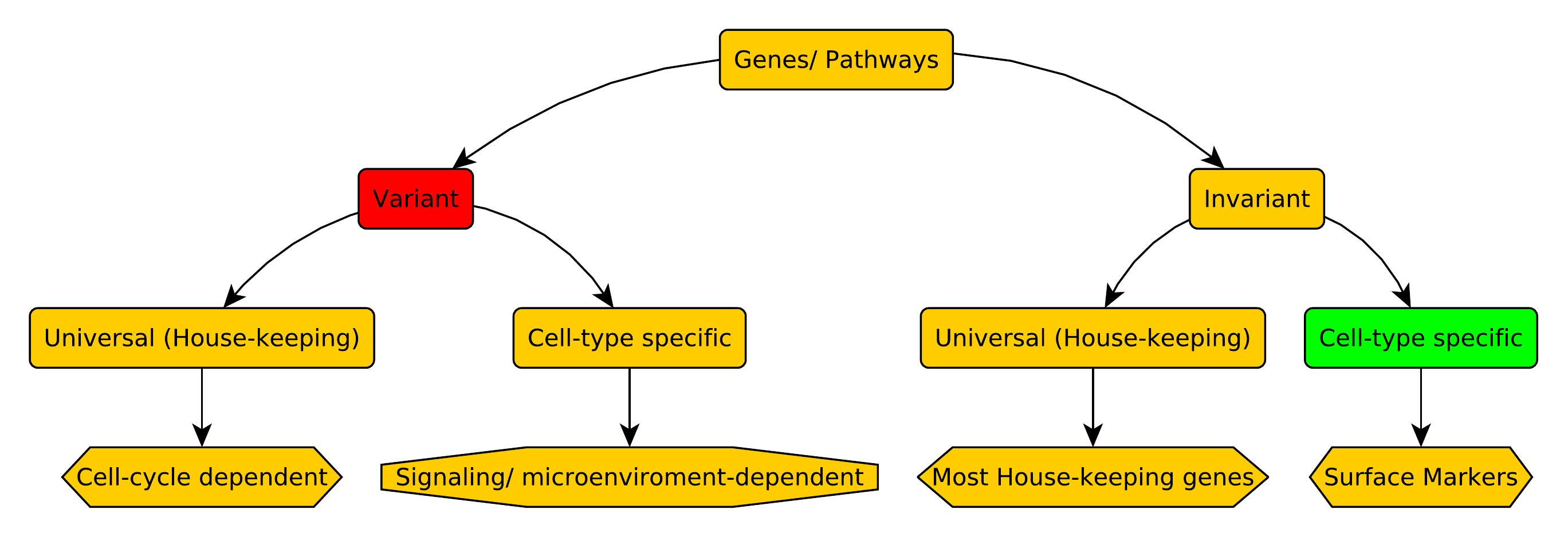}
\caption{High-level classification of genes}
\label{fig:gene_classification}
\end{figure}

We note that there is no prior knowledge as to the ``proper" choice of the marker selection method 
in the literature and that their effect on the deconvolution performance is 
unclear. An in-depth comparison of marker detection methods can benefit future 
developments in this field. An ideal marker should serve two purpose: (i) be highly informative of 
the cell-type in which it is expressed, (ii) shows low variance due to spatiotemporal changes 
in the environment (changes in time or microenvironment). Figure~\ref{fig:gene_classification} 
shows a high-level classification of genes. An ideal marker is an invariant, cell-type 
specific gene, marked with green in the diagram. On the other hand, variant genes, both 
universally expressed and tissue-specific, are not good candidates, especially in cases 
where references are adopted from a different study. These genes, however, comprise ideal 
subsets of genes that should be updated in full deconvolution while updating matrix 
$\Matrix{G}$, since their expression in the reference profile may differ significantly from the true 
cell-types in the mixture. It is worth mentioning that the proper ordering to identify best 
markers is to first identify tissue-specific genes and then prune them based on their 
variability. Otherwise, when selecting invariant genes, we may select many housekeeping 
genes, since their expression is known to be more uniform compared to tissue-specific genes.

Another observation relates to the case in which groups of profiles of cell-types have high
similarity within the group, but are significantly distant from the rest. This makes 
identifying marker genes more challenging for these groups of cell-types. An instance of this 
problem is when we consider markers in the \textbf{PERT} datasets. In this case, erythrocytes have 
a much larger number of distinguishing markers compared to other references. This phenomenon is 
primarily attributed to the underlying similarity between undifferentiated cell-types in the 
\textbf{PERT} datasets, and their distance from the fully differentiated red blood cells. In 
these cases, it is beneficial to summarize each group of similar tissues using a ``representative
profile" for the whole group, and to use a hierarchical structure to recursively identify markers 
at different levels of resolution \cite{Mohammadi2015}.

Finally, we examine the common choice of condition number as the optimal choice to 
select the number of markers. First, unlike the ``U" shape plot reported in previous studies, 
in which condition number initially decreases to an optimal point and then starts 
increasing, we observe variable behavior in the condition number plot, both for Newman 
\textit{et al.} and Abbas \textit{et al.} methods. This makes the generalization of condition 
number as a measure applicable to all datasets infeasible. Additionally, we note that the 
lowest condition number is achieved if $\Matrix{G}$ is fully orthogonal, that is 
$\Matrix{G}^{T}\Matrix{G} = \kappa \Matrix{I}$ for any constant $\kappa$. By selecting 
tissue-selective markers, we can ensure that the product of columns in the resulting matrix is 
close to zero. However, the norm-2 of each column can still be different. We developed a 
method that specifically grows the basis 
matrix by accounting for the norm equality across columns. We find that in all cases our 
basis matrix has a lower condition number than both the Newman \textit{et al.} and Abbas \textit{et 
al.} methods, but it did not always improve the overall performance of deconvolution methods 
using different loss functions. Further study on the optimal choice of the number of markers 
is another key question that needs further investigation

\subsection{To Regularize or Not to Regularize}
\label{sec:regularization_effect}

We now evaluate the impact of regularization on the performance of different deconvolution 
methods. To isolate the effect of the regularizer from prior filtering/feature selection steps, we 
apply regularization on the original datasets. The $\Reg_1$ 
regularizer is typically applied in cases where the solution space is large, that is, the total number 
of available reference cell-types is a superset of the true cell-types in the mixture. This type of 
regularization acts as a ``selector" to choose the most relevant cell-types and zero-out the 
coefficients for the rest of the cell-types. This has the effect of enforcing a sparsity pattern. 
Datasets used in this study are all controlled benchmarks in which references are 
hand-picked to match the ones in the mixture; thus, sparsifying the solution does not add value 
to the deconvolution process. On the other hand, an $\Reg_2$ regularizer, also known as Tikhonov 
regularization, is most commonly used when the problem is ill-posed. This is the case, for example,
when the underlying cell-types are highly correlated with each other, which introduces dependency 
among columns of the basis matrix. In order to quantify the impact of this type of regularization 
on the performance of deconvolution methods, we perform an experiment similar to the one in 
Section~\ref{sec:loss_effect} with an added $\Reg_2$ regularizer. In this experiment, we use
$\Loss_1$ and $\Loss_2$ loss functions, as we previously showed that the performance of the
other two loss functions is similar to $\Loss_1$. Instead of using Ridge regression introduced
in Section~\ref{sec:Ridge}, we implement an equivalent formulation,
$\parallel \Vector{m} - \Matrix{G}\Vector{c}\parallel_2 + \lambda \parallel \Vector{c} \parallel_1$,
which traces the same path but has higher numerical accuracy. To identify the optimal value of the
$\lambda$ parameter that balances the relative importance of solution fit versus regularization,
we search over the range of $\{10^{-7}, \cdots, 10^7\}$. It is notable here that when $\lambda$
is close to zero, the solution is identical to the one without regularization, whereas when
$\lambda \rightarrow \infty$ the deconvolution process is only guided by the solution size.
Similar to the range filtering step in Section~\ref{sec:range_filtering_effect}, we use the
minimum \textit{mAD} error to choose the optimal value of $\lambda$.

\begin{figure}
\centering
\includegraphics[width=0.9\columnwidth]{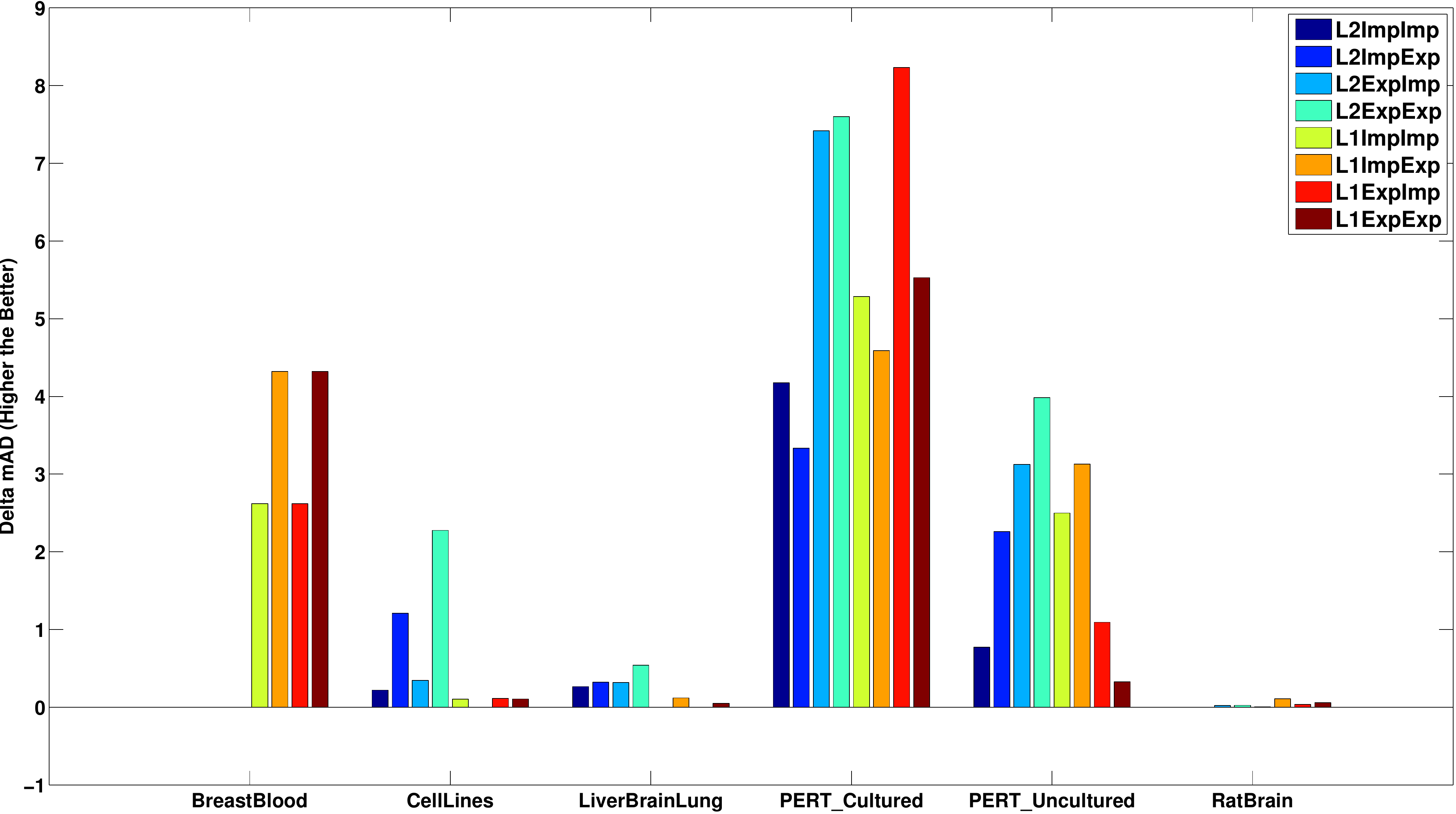}
\caption{Effect of L2 regularization on the performance of deconvolution methods}
\label{fig:reg_delta_mAD}
\end{figure}

Figure~\ref{fig:reg_delta_mAD} presents changes in mAD error,
compared to original errors, after regularizing loss functions with the $\Reg_2$ regularizer.
From these observations, it appears that \textbf{PERT\_Cultured} has the most gain due to
regularization, whereas for \textbf{PERT\_Uncultured}, the changes are smaller.
A detailed investigation, however, suggests that in the majority of cases for
\textbf{PERT\_Cultured}, the performance gain is due to over shrinkage of vector
$\Vector{c}$ to the case of being almost uniform. Interestingly, the choice of
uniform $\Vector{c}$ has lower \textit{mAD} error for this dataset compared to most
other results. Overall, both of the \textbf{PERT} datasets show significant 
improvements compared to the original solution, which can be attributed to the underlying
similarity among hematopoietic cells. On the other hand, an unexpected observation is the
performance gain over $\Loss_1$ configurations for the \textbf{BreastBlood} dataset. This is primarily
explained by the limited number of cell-types (only two), combined with the similar
concentrations used in all samples (only combinations of $67\%$ and $33\%$).

\begin{figure}
\centering
\includegraphics[width=0.9\columnwidth]{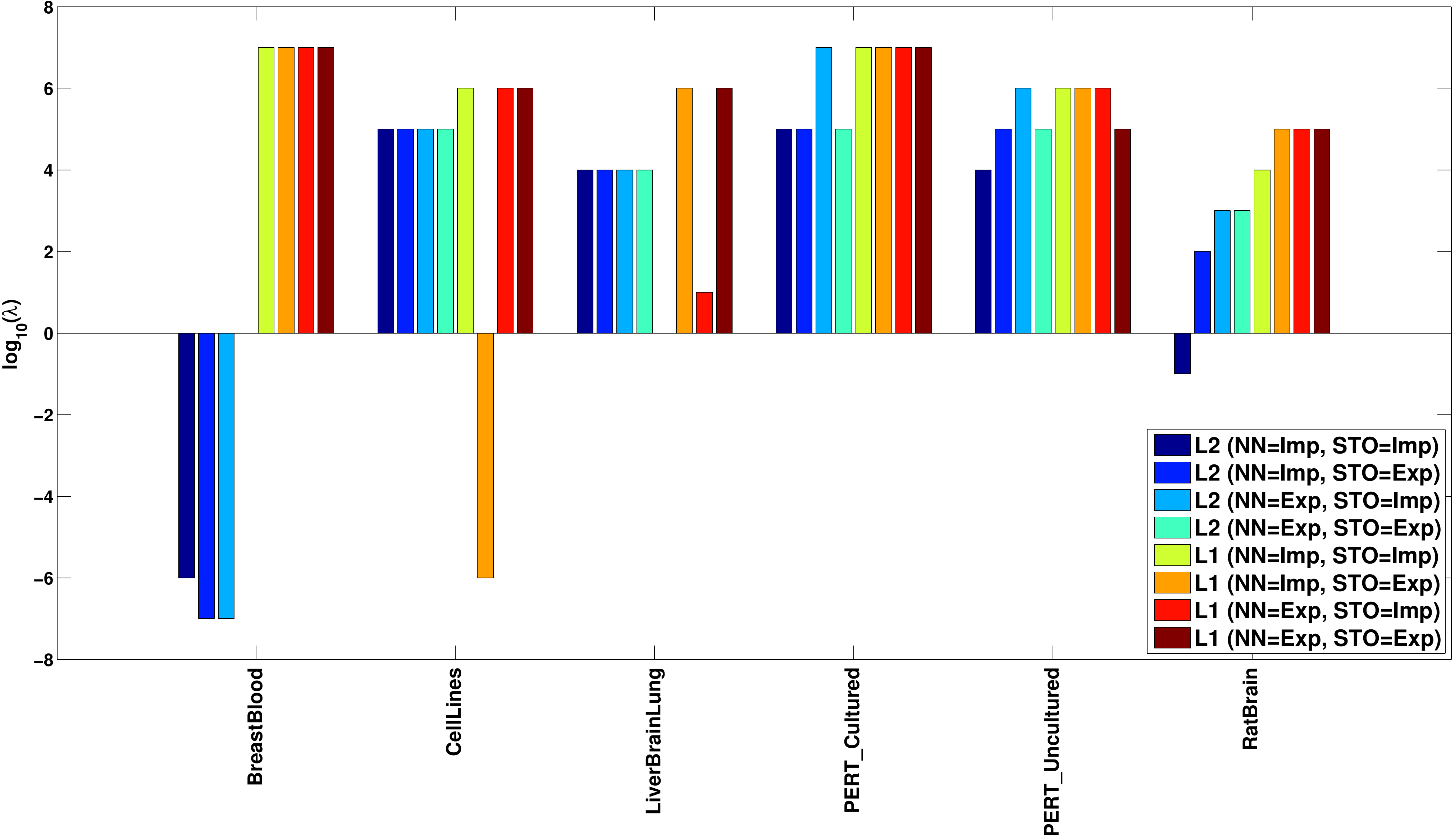}
\caption{Optimal value of $\lambda$ for each dataset/configuration pair}
\label{fig:reg_lambda_choice}
\end{figure}

To gain additional insight into the parameters used in each case during deconvolution, we plot the
optimal $\lambda$ values for each configuration in each dataset. Figure~\ref{fig:reg_lambda_choice} 
summarizes the optimal values of the $\lambda$ parameter. Large values indicate a beneficial effect 
for regularization, whereas small values are suggestive of negative impact. In all cases where 
the overall mAD score has been improved, their corresponding $\lambda$ parameter was large. However, 
large values of $\lambda$ do not necessarily indicate a significant impact on the final solution, 
as is evident in the \textbf{CellLines} and \textbf{LiverBrainLung} datasets. Finally, we 
observe that cases where the value of $\lambda$ is close to zero are primarily associated with
the $\Loss_2$ loss function.

\subsection{Summary}
\label{sec:summary}

Based on our observations, we propose the following guidelines for the
deconvolution of expression datasets:

\begin{enumerate}
\item
Pre-process reference profiles and mixtures using invariant, universally expressed
(housekeeping) genes to ensure that the \textit{similar cell quantity (SCQ)} constraint is satisfied.

\item Filter violating features that cannot satisfy the \textit{sum-to-one (STO)} constraint.

\item Filter lower and upper bounds of gene expressions using adaptive range filtering.

\item Select invariant (among references and between references and samples) cell-type-specific
markers to enhance the discriminating power of the basis matrix.

\item Solve the regression using the $\Loss_1$ loss function with explicit constraints (check),
together with an $\Reg_2$ regularizer, or group LASSO if sparsity is desired among groups of
tissues/cell-types.

\item Use the  L-curve method to identify the optimal balance between the regression fit and the
regularization penalty.
\end{enumerate}

\section{Concluding Remarks}
\label{sec:conclusion}

In this paper, we present a comprehensive review of different methods for deconvolving 
linear mixtures of cell-types in complex tissues. We perform a systematic analysis of the
impact of different algorithmic choices on the performance of the deconvolution methods,
including the choice of the loss function, constraints on solutions, data filtering,
feature selection, and regularization. We find $\Loss_2$ loss to be superior in cases where
the reference cell-types are representative of constitutive cell-types in the mixture,
while $\Loss_1 $ outperforms the $\Loss_2$ in cases where this condition does not hold.
Explicit enforcement of the sum-to-one (STO) constraint typically degrades the performance of
deconvolution. We propose simple bounds to identify features violating this constraint and evaluate
the total number of violating features in each dataset. We observe an unexpectedly high number of
features that cannot satisfy the STO condition, which can be attributed to problems with normalization
of expression profiles, specifically normalizing references and samples with respect to each 
other. In terms of filtering the range of expression values, we find that fixed thresholding is 
not effective and develop an adaptive method for filtering each dataset individually. 
Furthermore, we observed that range filtering is not always beneficial for deconvolution and,
in fact, in some cases it can deteriorate the performance. We implement two commonly used
marker selection methods from the literature to assess their effect on the 
deconvolution process. Orthogonalizing reference profiles can enhance the discriminating power 
of the basis matrix. However, due to known correlation between the mean and variance of 
expression values, this process alone does not always provide satisfactory results. Another 
key factor to consider is the low biological variance of genes in order to enhance the 
reproducibility of the results and allow deconvolution with noisy references.
The combination of range filtering and marker selection eliminates genes 
with high mean expression, which in turn enhances the observed results. Finally, we address
the application of Tikhonov regularization in cases where reference cell-types are highly 
correlated and the regression problem is ill-posed.

We summarize our findings in a simple set of guidelines and identify open problems
that need further investigation. Areas of particular interest for future research include:
(i) identifying the proper set of filters based on the datasets, (ii) expanding deconvolution
problem to cases with more complex, hierarchical structure among reference vectors, and
(iii) selecting optimal features to reduce computation time while maximizing the
discriminating power.

\section*{Acknowledgment}

This work is supported by the Center for Science of Information (CSoI), an NSF Science and 
Technology Center, under grant agreement CCF-0939370, and by NSF Grant BIO 1124962.

\ifCLASSOPTIONcaptionsoff
  \newpage
\fi



\printbibliography

\begin{IEEEbiography}[{\includegraphics[width=1in,height
=1.25in,clip,keepaspectratio]{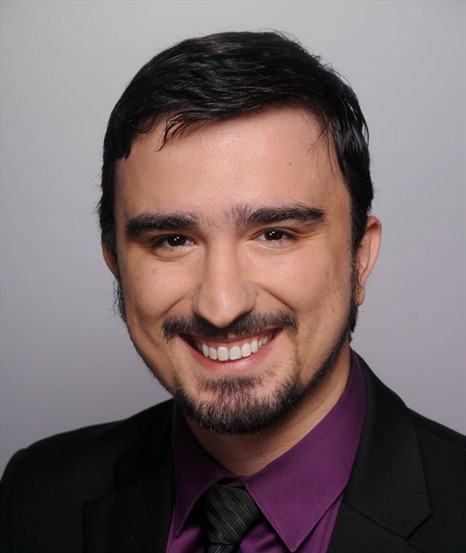}}]{Shahin Mohammadi}  received his Master's degree in Computer Science from Purdue University in Dec. 2012 and is currently a Ph.D. candidate at Purdue. His research interests include computational biology, machine learning, and parallel computing. His current work spans different areas of Bioinformatics/Systems Biology and aims to develop computational methods coupled with statistical models for data-intensive problems, with application in mining the human tissue-specific transcriptome and interactome.
\end{IEEEbiography}

\begin{IEEEbiography}[{\includegraphics[width=1in,height
=1.25in,clip,keepaspectratio]{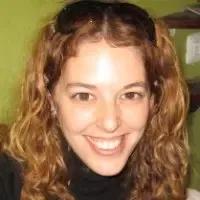}}]{Neta Zuckerman}  received her PhD degree in computational biology from the University of Bar-Ilan, Israel in June 2010. She has completed her post-doctorate in 2015 as a computational biologist at Stanford University, School of Medicine and City of Hope as well as a visiting scholar at the department of Electrical Engineering, Stanford University. Her research interests focus on investigating the role of immune cells in the setting of various diseases, specifically cancer, utilizing algorithm development and microarray data analysis. Neta is currently a computational biologist at Genentech Inc.
\end{IEEEbiography}

\begin{IEEEbiography}[{\includegraphics[width=1in,height
=1.25in,clip,keepaspectratio]{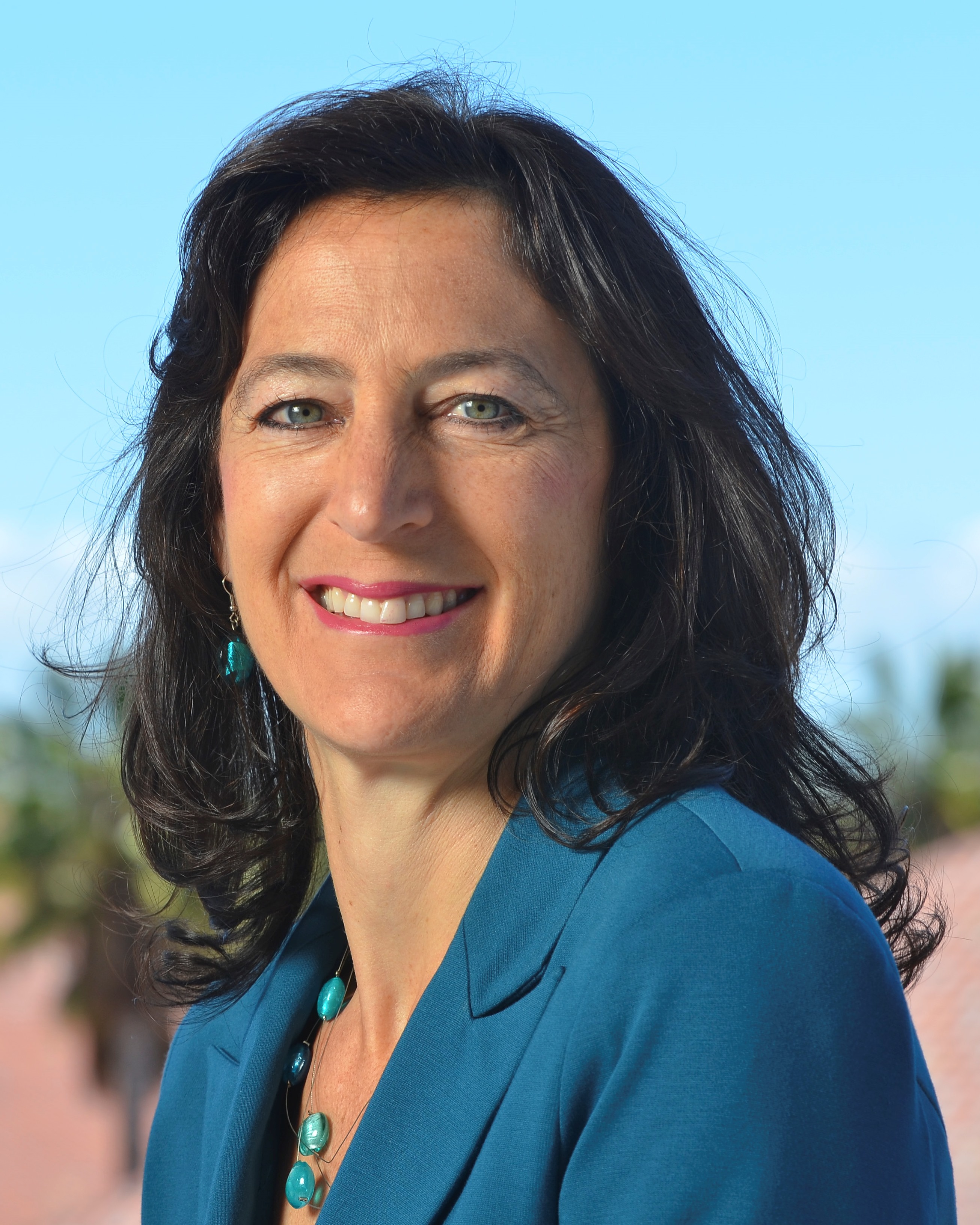}}]{Andrea Goldsmith}  Andrea Goldsmith is the Stephen Harris professor in the School of Engineering and a professor of Electrical Engineering at Stanford University. Her research interests are in information theory and communication theory, and their application to wireless communications as well as biology and neuroscience. She has received several awards for her work including the IEEE ComSoc Edwin H. Armstrong Achievement Award, the IEEE ComSoc and Information Theory Society Joint Paper Award, and the National Academy of Engineering Gilbreth Lecture Award. She is author of 3 textbooks, including Wireless Communications, all published by Cambridge University Press, as well as an inventor on 28 patents.
\end{IEEEbiography}

\begin{IEEEbiography}[{\includegraphics[width=1in,height
=1.25in,clip,keepaspectratio]{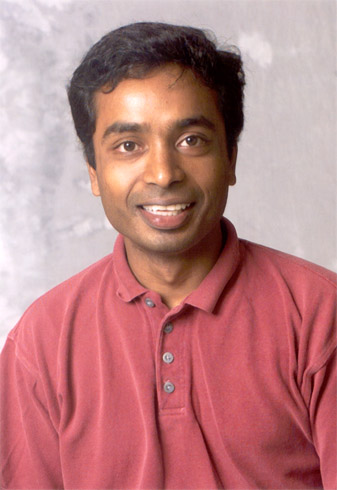}}]{Ananth Grama} received
the PhD degree from the University of Minnesota in 1996. He is
currently a Professor of Computer Sciences and Associate Director of
the Center for Science of Information at Purdue University.
His research interests span areas of
parallel and distributed computing architectures,
algorithms, and applications. On these topics, he
has authored several papers and texts. He is a member of the
American Association for Advancement of Sciences.
\end{IEEEbiography}

\end{document}